\documentclass[preprint,showpacs,preprintnumbers,amsmath,amssymb,aps]{revtex4-1}
\usepackage{times}
\usepackage{graphicx,dcolumn,bm}
\usepackage{amssymb,amstext,amsmath}
\usepackage{graphicx}
\usepackage{dcolumn}
\usepackage{bm}
\usepackage{amssymb}
\usepackage{multirow}
\usepackage{bigstrut}
\usepackage{makecell,rotating}
\usepackage{mathrsfs}
\usepackage{booktabs}
\usepackage{threeparttable}
\usepackage{multirow}
\usepackage{subfigure}
\usepackage{epsfig}
\usepackage{threeparttable}
\usepackage{chngpage}
\usepackage{float}
\usepackage{amstext}
\usepackage{amsmath}
\usepackage{setspace}

\usepackage[colorlinks=false,hyperindex,plainpages=true,bookmarks, pdfborder = 0 0 0]{hyperref}
\usepackage{hyperref}

\usepackage{color}
	 \definecolor{darkred}{rgb}{0.75,0,0}
	 \definecolor{darkgreen}{rgb}{0,0.5,0}
	 \definecolor{darkblue}{rgb}{0,0,0}
  	 \definecolor{darkorange}{rgb}{1,0.9,0.1}

\usepackage[top=23.4mm, bottom=23.4mm, left=23.4mm, right=23.4mm]{geometry}

\newcommand{\x}{\mathbf{x}}
\newcommand{\C}{\mathbf{c}}
\newcommand{\A}{\mathbf{A}}

\newcommand{\B}{\mathbf{B}}

\newcommand{\p}{\mathbf{D}}
\newcommand{\f}{\mathbf{F}}
\newcommand{\W}{\mathbf{W}}

\newcommand{\Weff}{\mathbf{W}_{\textrm{eff}}}
\newcommand{\zero}{\mathbf{0}}

\newcommand{\E}{E}

\renewcommand{\u}{\mathbf{u}}
\renewcommand{\d}{\mathbf{d}}
\newcommand{\s}{\mathbf{S}}
\newcommand{\e}{\mathrm{e}}

\newcommand{\ie}{\textit{i.e.}}

\renewcommand{\thefootnote}{\Alph{footnote}}

\newcommand{\Eub}{\overline{\E}}
\newcommand{\Elb}{\underline{\E}}

\begin{document} 

\title{Control energy scaling in temporal networks} 

\author{Aming Li,$^{1,2}$ Sean P. Cornelius,$^{1,3}$ Yang-Yu Liu,$^{3,4}$ Long Wang,$^{2,\ast}$ Albert-L\'{a}szl\'{o} Barab\'{a}si$^{1,4,5,6,}$}
\email{Corresponding authors.}
\affiliation{$^{1}$Center for Complex Network Research and Department of Physics,
Northeastern University, Boston, MA 02115, USA\\
$^{2}$Center for Systems and Control, College of Engineering,
Peking University, Beijing 100871, China\\
$^{3}$Channing Division of Network Medicine, Brigham and Women's Hospital,
Harvard Medical School, Boston, MA 02115, USA\\
$^{4}$Center for Cancer Systems Biology, Dana-Farber Cancer Institute,
Boston, MA 02115, USA\\
$^{5}$Department of Medicine, Brigham and Women's Hospital, Harvard Medical School,
Boston, MA 02115, USA\\
$^{6}$Center for Network Science, Central European University, Budapest 1052, Hungary
}

\date{\today}

%  \item Corresponding author. E-mail: longwang@pku.edu.cn (L.W.), alb@neu.edu (A.-L.B.).

\begin{abstract}
%Tradional studies of controlcomplex networks have focused on static versions where the embedded topology constructions are time-invariant.
%Here we develop a framework to model temporal networks as switched systems in control theory.
%The controllable space of temporal networks is obtained, which could tell when the temporal network is controllable.
%From human interactions data, we find that the time that temporal networks needed to be fully controllable is always less than that for aggregated static networks, suggesting that the direct aggregation of a controllable temporal network always gives an uncontrollable static network.
%When randomization takes place on the data, the time decreases due to the ubiquitous bursts behavior in human interactions which impede the frequent switching of temporal networks and the change of the corresponding aggregated networks, thus elimination of bursts by randomizing could decrease the time.
%The optimal trajectory realizing the control of temporal networks from one state to another is also given, along which the control energy is minimum.
%For temporal networks, the control energy is less than that for aggregated when the networks switch quickly.
%Surprisingly, we find that the scaling behavior of energy for controlling a temporal network with a series of snapshots is determined by the first and the last snapshot.
In practical terms, controlling a network requires manipulating a large number of nodes with a comparatively small number of external inputs, a process that is facilitated by paths that broadcast the influence of the (directly-controlled) driver nodes to the rest of the network.  
%For temporal networks, although paths herein are seldom instantaneously available, it has been shown that the evolution of network topology saves orders of magnitude control energy compared to the canonical static networks.
Recent work has shown that surprisingly, temporal networks can enjoy tremendous control advantages over their static counterparts despite the fact that in temporal networks such paths are seldom instantaneously available.
To understand the underlying reasons, here we systematically analyze the scaling behavior of a key control cost for temporal networks---the control energy. 
We show that the energy costs of controlling temporal networks are  determined solely by the spectral properties of an ``effective" Gramian matrix, analogous to the static network case.
%, analogous to the static network case.
%and find that, there are simply determined by the maximum and minimum eigenvalues of the effective controllability Gramian matrix.
Surprisingly, we find that this scaling is largely dictated by the first and the last network snapshot in the temporal sequence, independent of the number of intervening snapshots, the initial and final states, and the number of driver nodes. 
%All theoretical results are validated by numerical calculations.
Our results uncover the intrinsic laws governing why and when temporal networks save considerable control energy over their static counterparts.
\end{abstract}

\maketitle

\newpage
\tableofcontents

\newpage
\section{Introduction}
%Complex systems could be investigated from the perspective of their underlying complex networks skeleton, where nodes indicate units of the systems and edges capture who interacts with whom, while an ultimate goal of the investigations is trying to control them.
%An ultimate goal of exploring complex systems is trying to actualize the control of them after acquiring their embedded topological constructions where nodes indicate units of the systems and edges capture who interacts with whom.
%Indeed, implementations of the control signals could drive a system from arbitrary states to the desired, based on which we could steer the system to function along our expectations.
A central goal in many applications of networked systems is the control of network dynamics. 
Indeed, problems as diverse as power system stability \cite{Buldyrev2010NatCascade}, cell reprogramming \cite{ConRealSysCell11}, and maintenance of gut microbiome health \cite{Nat2012Microbiota,Coyte2015Sci} all require the ability to steer a system to (or keep it in) a desirable state.
%Indeed, problems as diverse as microbial systems stability, cell reprogramming, and power grid management all require the ability to steer a system to function according to expectations using external influence \cite{Nat2012Microbiota,Coyte2015Sci,Buldyrev2010NatCascade,ConRealSysCell11}.
Based on the idea of structural controllability from control theory \cite{Lin1970}, Liu \textit{et al.}  devised an efficient algorithm to determine the minimal number of nodes required to control complex networks with a particular class of dynamics \cite{Liu2011}.
And in the past several years, numerous subsequent investigations have emerged focusing on problems as diverse as
%\cite{Pequito2013,Olshevsky2014}, 
classification of control nodes \cite{Jia2013}, control profiles \cite{Ruths2014science}, target control \cite{Gao2014}, control of edge dynamics \cite{Nepusz2012}, and also the energy (or cost) required for control in practice \cite{Yan2012PRL,energy2014,Yan2015a}. 

Yet most existing studies of controllability have been premised on \emph{static} networks \cite{Yan2012PRL,Nepusz2012,Jia2013,Sun2013prl,Wang2013,Ruths2014science,Chen2014,Gao2014,energy2014,Yan2015a,Chen2017}, with comparatively limited attention devoted to the case of (discrete-time) dynamics on temporal networks \cite{Posfai2014NJP,LiXiang2014}. 
Putatively static networks are often aggregated from an underlying temporal sequence of snapshots, representing subsets of nodal interactions active at any given time.
With this recognition that temporal networks are in many areas the rule rather than the exception, many studies have explored temporal analogues of important structural features of static networks including the
small-world \cite{Watts1998a} and scale-free \cite{Barabasi1999a} properties, and community structure \cite{Girvan02PNASCommunity}. 
%have been discovered successively, and various dynamical processes are also explored over static networks \cite{barrat2008dynamical,Arenas2008PR,Pastor2001aEpideNoThers,Baruch13NatPhy,Santos05PRL,Li2015JTB}.
But the temporal nature of networks cannot be neglected for many \emph{dynamical} processes on networks either \cite{barrat2008dynamical,Arenas2008PR,Pastor2001aEpideNoThers,Baruch13NatPhy,Santos05PRL,Li2015JTB,PercoDynaNet2010}. 
Indeed, consider that if Alice interacts with Charlie after first interacting with Bob, then information (or a virus) cannot be propagated from Charlie to Bob  through Alice.
%due to the time order on the temporal (time variant) network.
The effects of such timing constraints on system dynamics have been reported on accessibility \cite{Accessibility13PRL},  diffusion or epidemic spreading \cite{Slowingdon12PRE,Slowingd13PRL,Ribeiro13SR,Causality14NatCom}, and human cooperative behavior on dynamical population structures \cite{LiGameTemp}.

Recent research has revealed that control, too, is a dynamical process profoundly affected by network temporality, and in a surprising way \cite{Li2016}.
It has been shown that temporal networks enjoy control costs orders of magnitude lower compared to their static counterparts \cite{Li2016}. 
Yet, the laws governing the control costs for temporal networks remain elusive.
Here we focus on the behavior of one key control cost---the control energy---to control temporal networks, deriving a simple rule that governs the scaling of the control energy with the dynamical evolution of the network topology.

\section{Control Energy}

We regard a temporal network as an ordered sequence of $M$ separate networks called \emph{snapshots} on a fixed set of $N$ nodes, and we denote by $\A_m$ the adjacency matrix of snapshot $m$ for $m=1,2,\cdots,M$. 
Starting from the first snapshot at time $t_0$, we assume each snapshot $m$ lasts for a duration of $\tau_m$ time units. 
We consider networks whose dynamical state follows
\begin{equation}
\label{temporaldynamics}
\dot{\x}(t)=\A_m \x(t) + \B \u(t)
\end{equation}
over the time interval $t \in [t_{m-1}, t_{m-1}+\tau_m)$, where $t_{m-1}=\sum_{j=1}^{j=m-1}\tau_j$ and $x_i(t)$ is the state of node $i$ at time $t$ with $\x(t) = (x_1(t),x_2(t),\cdots,x_N(t))^\textrm{T} \in\mathbb{R}^N$. 
Here, $\u_m(t)\in\mathbb{R}^p$ is a vector containing the $p$ independent control inputs and $\B$ gives the (constant) mapping between these inputs and the driver nodes of the network--those that receive input directly. 
%We will typically consider the case that one input corresponds to exactly one driver node.
We will focus on the case where one input corresponds to one driver node, as has been the norm in previous studies of network control \cite{Yan2012PRL,energy2014,Yan2015a,Liu2016Rev,Klickstein2017}.

The canonical definition of the control energy 
required to drive a system from state $\x_0$ at $t_0$ to $\x_\text{f}$ at $t_f$ 
is $\frac{1}{2}\int_{t_0}^{t_f}\u^{\textrm{T}}(t)\u(t)\textrm{d}t$, a definition that applies to arbitrary systems, whether linear or nonlinear, temporal or static \cite{Rajapakse2011pnas,OptimalBooLewis,Yan2012PRL,Sun2013prl}.
In the case of a temporal network obeying Eq.~(\ref{temporaldynamics}), we have shown \cite{Li2016} that the corresponding energy-optimal control signal  can be constructed piecewise as $\u(t)=\B^\textrm{T}\e^{\A^\textrm{T}_{m}(t_{m}-t)}\C_{m}$ for $t \in [t_{m-1}, t_{m})$.
This signal is parameterized by the constant vectors $\C_m$ $=$ $(c_{m,1},$ $c_{m,2},$ $\cdots,$ $c_{m,N})^{\textrm{T}}$ $\in$ $\mathbb{R} ^N$, which are unique and can be calculated according to the quadratic programming problem:
\begin{eqnarray}
\label{quadprog}
  \textrm{min} & & ~~~~E(\x_0, \x_\text{f}) =
  \frac{1}{2}\C^\textrm{T}\W\C \nonumber \\
  \textrm{s.t.} & &~~~~ \textrm{\textbf{SWc}}=\d
\end{eqnarray}
where 
$\C = 
 \left(\C^{\textrm{T}}_{1},\C^{\textrm{T}}_{2},\cdots,\C^{\textrm{T}}_{M} \right) ^ {\textrm{T}}
$,
$\s = \left(  \prod_{l=M}^2 \textrm{e}^{\A_{l}\tau_l}, \cdots,  \prod_{l=M}^{m+1} \textrm{e}^{\A_{l}\tau_l}  ,\cdots,\textrm{\textbf{I}}_N \right)$.
$\W =$
$\textrm{diag}(\W_{1},$
$\cdots,\W_{m},$
$\cdots,$
$\W_{M})$ is block-block diagonal, containing the controllability Gramians of each of the snapshots viewed as isolated systems, \ie, 
$\W_{m}
= 
\int_{t_{m-1}}^{t_m}\textrm{e}^{\A_{m}(t_m-s)}\B\B^\textrm{T}
\textrm{e}^{\A^\textrm{T}_{m}(t_{m}-s)}\mathrm{d}s
$.
We denote by $\d$ the difference between the desired final state $\x_\text{f}$ and the state that the system would reach naturally from $\x_0$ in the absence of control, namely
$\d
= 
\x_\text{f} - \prod_{l=M}^1 \textrm{e}^{\A_{l}\tau_l} \x_0$.
%(detailed derivation process is illustrated in SI \ref{quadraticprogramming}).
In plain English, this problem seeks the minimum control energy while satisfying that the initial state be $\x_0$, the final state $\x_\text{f}$, and the end state in any given snapshot is the initial state of the next.

\section{Bounds of the Optimal Control Energy}
%After getting the optimal solution of (\ref{quadprog}) %(see SI)
%we have that the optimal input energy required to control a temporal network is 
One can solve (\ref{quadprog}) analytically and find that the minimal energy required to control a temporal network between initial state $\x_0$ and final state $\x_\text{f}$ is
\begin{equation}
\label{retemp}
E^* (\x_0, \x_\text{f})
 = 
\frac{1}{2}\d^\textrm{T} \Weff^{-1}\d,
\end{equation}
where $\Weff = \s\W\s^\textrm{T}$.
%$\C^*  =  \s^\textrm{T} \Weff ^{-1}\d$, and $\d$ is the vector between the desired final state $\x_\text{f}$ and the natural final state that the system will reach without control inputs.
For a given pair of initial and final states, the control energy is thus determined by the spectral properties of the ``effective'' Gramian matrix $\Weff$, analogous to the static network \cite{Yan2012PRL}. 
Henceforth, we will focus on the case $\x_0 = \zero$ (for the general initial states, please refer to the SI).
By normalizing so that $\x_\text{f}$ lies at unit distance we can consider the \emph{normalized} control energy,
 $E^* (\textbf{0}, \x_\text{f})  =  \x_\text{f}^\textrm{T}  \Weff  ^ {-1} \x_\text{f} / (2\x_\text{f}^\textrm{T}\x_\text{f})$.
Irrespective of the location of $\x_\text{f}$, this allows us to impose lower $\underline{E}$ and upper bounds $\overline{E}$ on the control energy as
$$\underline{E}=1/(2\eta_{\max}) \leq  E^* (\zero, \x_\text{f})  \leq  \overline{E} = 1/(2\eta_{\min}),$$
where $\eta_{\max}$ and $\eta_{\min}$  are the maximum and minimum eigenvalues of $\Weff$.
Since $\Weff$ is a real and symmetric matrix, all eigenvalues are real and the minimum and maximum are well-defined.
Note that when all snapshots are identical, meaning the network structure is time-invariant, $\Weff$ reduces to the typical controllability Gramian for static networks \cite{Yan2012PRL} (for a proof of this, please see Sec.~\ref{StatciCases} in the SI).
%As $\A_i$ is a constant matrix, our results for temporal networks could cover the static case \cite{Yan2012PRL}, where every snapshot in temporal network keeps the same, and derivations are given in SI (\ref{StatciCases}).
The above bounds apply to arbitrary temporal sequences, regardless of whether the dynamics of the constituent snapshots are stable, unstable, or a mix. 
This will allow us to systematically study the behavior of the control energy for a range of temporal networks and determine the regimes in which they have an advantage over their static counterparts.

\section{The Scaling Behavior of the Bounds for Two Snapshots}

The lower (upper) bound $\underline{E}$ ($\overline{E}$) of the optimal control energy indicates the best (worst) case control direction, that is, the direction of the eigenvector corresponding to the maximum (minimum) eigenvalue $\eta_\textrm{max}$ ($\eta_\textrm{min}$) of $\Weff$.
The properties of the corresponding eigenvalues in turn determine the scaling behavior of
%($\underline{E} = 1/(2\lambda_\textrm{max})$ is the control energy $E^*$ for $\x_\text{f}$ along )
%We find that the lower bound first decreases at scaling $\underline{E} \sim \Delta t ^{-1}$ for small $\Delta t$, and then keeps as a constant $-1/\sum_{c \in \mathcal{I}} \A_M^{-1}(c,c)$, where $\mathcal{I}$ is the set of driver nodes (see SI ?).
$\underline{E}$ and $\overline{E}$. 
%are determined by the maximum and minimum eigenvalues of the snapshots, respectively.
To understand the scaling behavior of $\underline{E}$ $(\overline{E})$ with respect to the duration time $h$ of each snapshot, we first analyze the case of  two snapshots $(\A_1, \B)$ and $(\A_2, \B)$, and later generalize to an arbitrary number of snapshots.
By approximating the maximum (minimum) eigenvalues $\lambda_{\max}^{(1)}$ $(\lambda_{\min}^{(1)})$  and $\lambda_{\max}^{(2)}$ $(\lambda_{\min}^{(2)})$ of $\A_1$ and $\A_2$ (see SI Sec.~\ref{Energyfor2netsReach}), we can obtain an analytic prediction of the scaling behavior of the $\underline{E}$ ($\overline{E}$) for controlling temporal networks from $\zero$ to $\x_\text{f}$.
% in Table~\ref{Tab_min_2}, and $\overline{E}$ is given in Table~\ref{Tab_max_2}.

\begin{figure}[H]
\centering
\includegraphics[width=\textwidth]{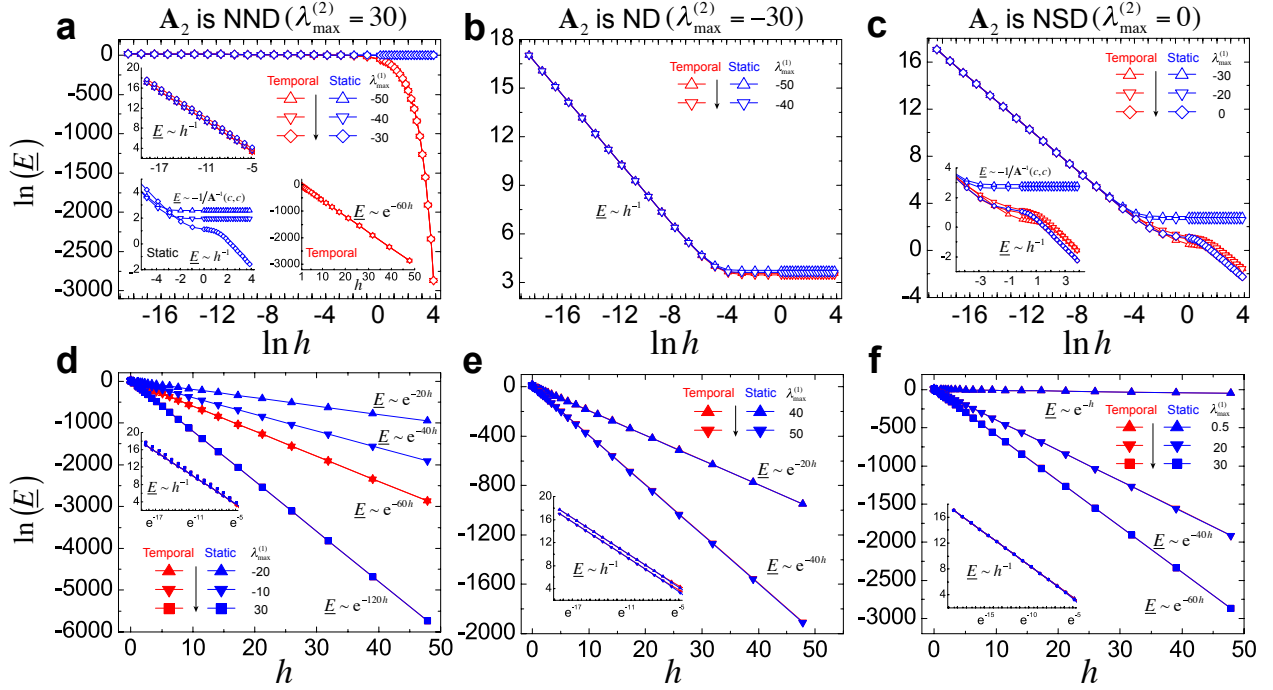}
\caption{
\textbf{Lower bound of the minimum energy needed for controlling temporal and static networks.}
Numerical results agree with the theoretical calculations shown in Table~\ref{Tab_min_2}, where each case is divided according the maximum eigenvalue $\lambda_{\max}^{(2)}$ of the second snapshot $\A_2$.
We set $\lambda_{\max}^{(2)} = 30$, $-30$, and $0$ to represent that $\A_2$ is NND (\textbf{a} and \textbf{d}), ND  (\textbf{b} and \textbf{e}), and NSD (\textbf{c} and \textbf{f}), respectively.
The black arrow in an inset of each panel indicates the difficulty level of controlling the networks from $\x_0 $ to $\x_\text{f}$ with $\| \x_\text{f} \| = 1$, along which less and less energy is required as $\lambda_{\max}^{(1)}$ increases.
$\underline{E} \sim h^{-1}$ when $h$ is small as shown in the insets of each panel except in (b), and then $\underline{E}$ decreases exponentially with the increase of maximum eigenvalue $\lambda_{\max}^{(1)}$ of the first snapshot $\A_1$ (d, e, and f).
When $\lambda_{\max}^{(2)} > 0$, $\underline{E}$ decreases exponentially for temporal network when $h$ is large, while staying constant for static network in the case where $\lambda_{\max}^{(1)} + \lambda_{\max}^{(2)} \leq 0$ (a).
All notation is the same as that in Table~\ref{Tab_min_2}.
The corresponding results for the case of more snapshots can be found in Figs.~\ref{figsi_min1} and \ref{figsi_min2}.
%For the Laplacian matrix, $a_{ij} \in (0, 1)$ with uniform distribution and maximum eigenvalue $a_1$ (see SI~\ref{LaplacianMatrix}).
All results correspond to a single representative network where $N = 20$, $k = 6$, $w_{ij} \in (0,1)$ uniformly, with a single node randomly chosen to receive the input signal.
 }
 \label{energy_2min}
\end{figure}

\begin{figure}[H]
\centering
\includegraphics[width=\textwidth]{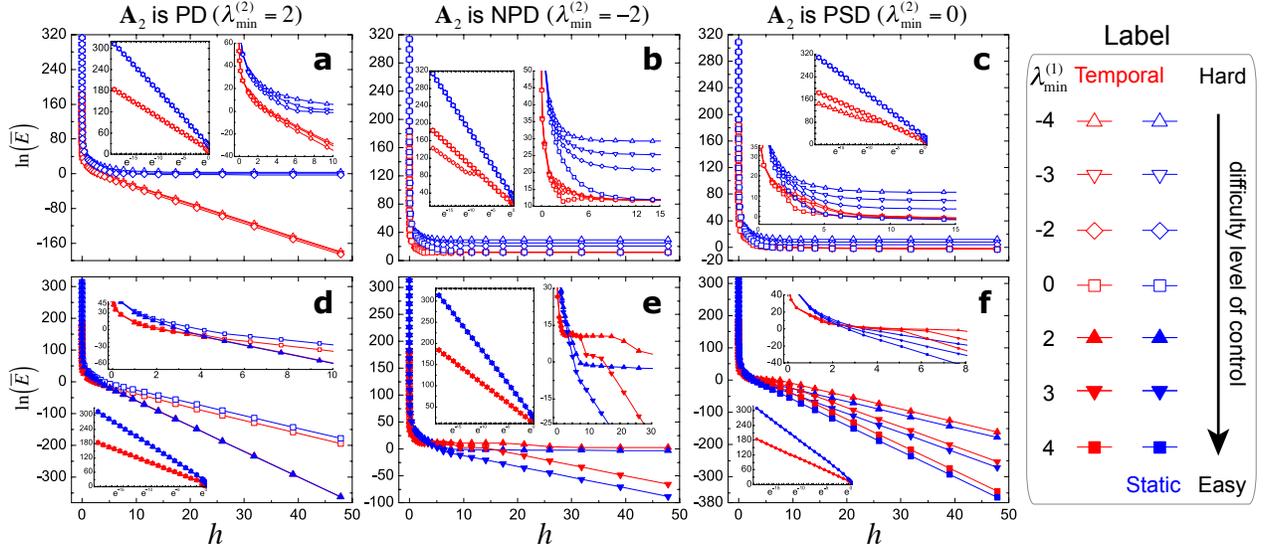}
\caption{
\textbf{Upper bound of the minimum energy needed for controlling temporal and static networks.}
Numerical results agree with the theoretical calculations shown in Table~\ref{Tab_max_2}, where each case can be divided according the minimum eigenvalue $\lambda_{\min}^{(2)}$ of the second snapshot $\A_2$.
We employ $\lambda_{\min}^{(2)} = 2$, $-2$, and $0$ to cover the cases in which $\A_2$ is PD (\textbf{a} and \textbf{d}), NPD (\textbf{b} and \textbf{e}), and PSD (\textbf{c} and \textbf{f}), respectively.
%Six panels correspond to that second snapshot is Positive Definite (PD, $\gamma_N > 0$, (\textbf{a}) and (\textbf{d})), Not Positive Definite (NPD, $\gamma_N < 0$, (\textbf{b}) and (\textbf{e})), and Positive Semi-Definite (PSD, $\gamma_N = 0$, (\textbf{c}) and (\textbf{f})).
%The arrow in each panel indicates the difficulty level of controlling the networks, which needs less energy from the initial state $\zero$ to the final state $\x_\text{f}$ with the increase of the minimum eigenvalue $\theta_N$ of the second snapshot $\A_2$.
%For the Laplacian matrix, $a_{ij} \in (-1, 0)$ with uniform distribution and minimum eigenvalue $a_1$ (see SI~\ref{LaplacianMatrix}).
When the duration time $h$ of each snapshot is short, the maximum energy for a temporal network is always less than that of its static counterpart (see the inset of each panel).
Furthermore, as $\lambda_{\min}^{(1)}$ increases, $\overline{E}$ decreases exponentially (second row).
 $\overline{E}$ always decreases exponentially for temporal network when $\lambda_{\min}^{(2)} > 0$ while static network keeps constant when $\lambda_{\min}^{(1)} + \lambda_{\min}^{(2)} \leq 0$ for large $h$ (see (\textrm{\textbf{a}})).
All notation is the same as that in Table~\ref{Tab_max_2}.
The corresponding results for the case of more number of snapshots can be found in Figs.~\ref{figsi_max1} and \ref{figsi_max2}.
All results correspond to a single representative network where $N = 8$, $k = 4$, $w_{ij} \in (-1,0)$ uniformly, and a single node was chosen randomly to receive the input signal.
%and the random node which is chosen to receive input directly is node $5$.
 }
 \label{energy_2max}
\end{figure}

Table~\ref{Tab_min_2} summarizes the possible behaviors of $\underline{E}$, which we find is dominated by the maximum eigenvalue $\lambda_{\max}^{(2)}$ of the second snapshot $\A_2$ for large $h$.
In this regime, we can therefore separate the behavior of $\underline{E}$ into three cases based on the sign of $\lambda_{\max}^{(2)}$.
%Indeed, $\lambda_{\max}^{(2)}$ divides the law of $\underline{E}$ into there cases. 
(i) When $\A_2$ is Not Negative Definite (NND) ($\lambda_{\max}^{(2)} > 0$), we find that $\underline{E}$ decreases exponentially with the exponent $2\left[\lambda_{\max}^{(1)}\textrm{H}(\lambda_{\max}^{(1)}) + \lambda_{\max}^{(2)}\right]$, where $\textrm{H}(x)$ is the Heaviside step function, 
%satisfying $\textrm{H}(x < 0) = 0$ and $\textrm{H}(x \geq 0) = 1$
and $\lambda_{\max}^{(1)}$ is the maximum eigenvalue of the first snapshot $\A_1$.
(ii) When $\A_2$ is Negative Definite (ND) ($\lambda_{\max}^{(2)} < 0$), $\underline{E}$ remains constant when $\lambda_{\max} = \lambda_{\max}^{(1)} + \lambda_{\max}^{(2)} \leq 0$, otherwise decreases exponentially with exponent $\lambda_{\max}$.
(iii) When $\A_2$ is Negative Semi-Definite (NSD) ($\lambda_{\max}^{(2)} = 0$), $\underline{E} \sim h^{-1}$ for $\lambda_{\max} \leq 0$, and otherwise decreases exponentially with the exponent $\lambda_{\max}$.
When $h$ is small, the law of unique with $\underline{E} \sim h^{-1}$. 
These analytical predictions, which are summarized in Table~\ref{Tab_min_2} and Table~\ref{Tab_max_2}, are corroborated by numerical results (Figs.~\ref{energy_2min} and~\ref{energy_2max}).

Here we have employed the Laplacian matrix with self-loops to represent the weighted undirected snapshot $\A_m$. 
This allows us to tune the values $\lambda_{\min}^{(m)}$ and $\lambda_{\max}^{(m)}$, and $w_{ii} = \lambda^{(m)}-\sum_{j=1,j \neq i}^{N}w_{ij}$ with $w_{ij}$ indicating the weight of the link between nodes $i$ and $j$.
When $w_{ij} >0$, we can set $\A_m$ to be any of NND, ND, or NSD simply by changing $\lambda^{(m)} =  \lambda^{(m)}_{\max}$.
And when $w_{ij} <0$, we can similarly change $\lambda^{(m)} =  \lambda^{(m)}_{\min}$ to tune $\A_m$ among Positive Definite (PD), Not Positive Definite (NPD), and Positive Semi-Definite (PSD).
For the corresponding static network, we have $\A = \sum_{m=1}^{M}\A_m\tau_m/\tau$ for a duration time $\tau=\sum_{m=1}^{M}\tau_m$, and its maximum (minimum) eigenvalue is $\sum_{m=1}^{M}  \lambda^{(m)}_{\max}\tau_m/\tau$ ($\sum_{m=1}^{M}  \lambda^{(m)}_{\min}\tau_m/\tau$).
And we assume all snapshots' durations are identical ($\tau_m=h$ for all snapshots $m$) for simplicity.
We have checked the robustness of our results for other settings of link weight.

\section{The First and Last Snapshots Determine the Scaling Behavior}

%For a temporal network with an arbitrary number of snapshots (say, $M$), the control energy can be accounted based on each snapshot with 
We can evaluate the contribution each snapshot makes to the overall control energy using
the following expression (see SI Sec.~\ref{si_firstandlastSub})
\begin{eqnarray}
E(\x_0,\x_\text{f}) &=& \frac{1}{2}
\sum_{i=1}^{i = M} \left(\x _{i}- \textrm{e}^{\A_{i}h}\x _{i-1} \right)^{\textrm{T}} \W_{i}^{-1} \left(\x _{i}- \textrm{e}^{\A_{i}h}\x _{i-1} \right).
\label{totalenergy}
 \end{eqnarray}
We find that when we control a system from arbitrary $\x_0$ to $\x_\text{f}$, it is the first and last snapshots that determine the scaling behavior of the control energy required (see SI Sec.~\ref{si_firstandlastSub}).
This somewhat surprising result can be understood by
\begin{eqnarray}
E^*(\x_0,\x_\text{f}) =
\frac{1}{2}
\left( \x_0^{\textrm{T}} \textrm{e}^{\A_{1}^{\textrm{T}}h} \W_{1}^{-1} \textrm{e}^{\A_{1}h} \x_0 -
\x_0^{\textrm{T}} \textrm{e}^{\A_{1}^{\textrm{T}}h} \W_{1}^{-1} \x_1
- \x_{M-1} ^{\textrm{T}} \textrm{e}^{\A_{M}^{\textrm{T}}h} \W_{M}^{-1} \x_\text{f} +
\x_{f} ^{\textrm{T}}  \W_{M}^{-1} \x_\text{f}
\right),
\end{eqnarray}
which indicates that $E^*(\x_0,\x_\text{f})$ is dominated by $\A_1$ and $\A_M$ for any kind of inputs (this equation is derived by minimizing Eq.~(\ref{totalenergy})).
Thus, although the whole sequence of snapshots influences the exact control signal $\u(t)$ and globally optimal trajectory $\x^*(t)$, it is only the first and last snapshots
that determine the corresponding control energy. 
This can be understood by the fact, that it is these snapshots from which 
 the temporal network must ``lift off" from $\x_0$ and ``land" at final state $\x_\text{f}$.
% , considering that the system states evolve along
%%Thus after $M$ snapshots, the final state $\x _f$ at time $t_M$ is
%\begin{equation}
%\label{solutionfortem}
%\x _f = \prod_{m=M}^1 \e^{\A _{m}\tau_m}\x _0 + \sum_{m=1}^{M-1} \prod_{j=M}^{m+1}\e^{\A _{j}\tau_j} 
%\W_{m}[t_{m-1}, t_m]
% + \W_{M}[t_{M-1}, t_M].
%\end{equation}
%\begin{eqnarray}
%\label{solutionfortem}
%\x _f &=& \prod_{m=M}^1 \e^{\A _{m}\tau_m}\x _0 + \sum_{m=1}^{M-1} \left(\prod_{j=M}^{m+1}\e^{\A _{j}\tau_j}\int_{t_{m-1}}^{t_m}\e^{\A _{m}(t_m-s)}\B _{m}\u_m(s)\mathrm{d}s\right) \\ \nonumber
%   & & + \int_{t_{M-1}}^{t_M}\e^{\A _{M}(t_M-s)}\B _{M}\u_M(s)\mathrm{d}s.
%\end{eqnarray}
%the numerical as well as analytical results are given in SI (\ref{EnergyforMnets}).
%Similar to a plane, the air route is designed based on the location of departure and destination airports, while the taking off and landing processes drive the plane to leave the departure airport and reach the destination eventually, and except these two processes, it cruises just according to predefined route. 

\section{The Scaling Behavior of the Optimal Control Energy for Arbitrary Number of Snapshots}
\begin{figure}[H]
\centering
\includegraphics[width=\textwidth]{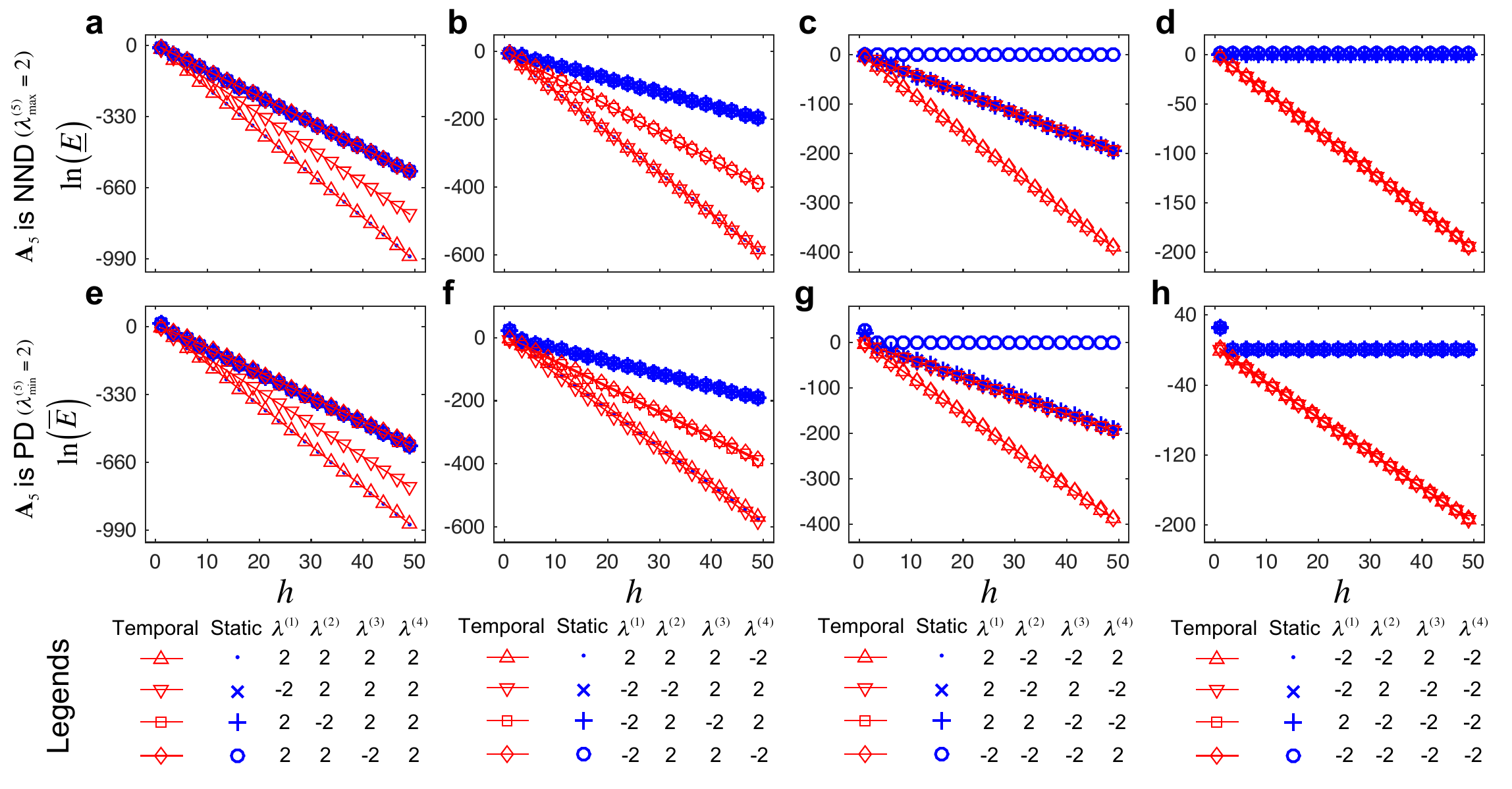}
\caption{
\textbf{Lower and upper bounds of the minimum energy needed for controlling temporal and static networks.}
For $M=5$ snapshots, we know that $\lambda_{\max}^{(5)}$ and $\lambda_{\min}^{(5)}$ of the last snapshot $\A_5$ dominate the scaling behavior of $\underline{E}$ and $\overline{E}$, respectively.
For example, when $\A_5$ is NND ($\lambda_{\max}^{(5)}=2$), $\underline{E}$ decreases exponentially with the exponent $\lambda_{\max} = \mathop {\max} \limits_{l} \left\{ 2 + \mathop {\sum_{m=l}^{M-1}} \limits_{1 \leq l \leq M-1}  \lambda_{\max}^{(m)} \right\}$.
From $\lambda_{\max}^{(m)} = \lambda^{(m)}$ in the legends, we show that numerical results agree with the theoretical calculations from \textbf{(a)} to \textbf{(d)}.
$\overline{E}$ decreases exponentially with the exponent 
$\lambda_{\min} = \mathop {\max} \limits_{l} \left\{ 2 + \mathop {\sum_{m=l}^{M-1}} \limits_{1 \leq l \leq M-1}  \lambda_{\min}^{(m)} \right\}$ when $\A_5$ is PD ($\lambda_{\min}^{(5)}=2$), and again we find the analytical results are validated by numerical calculations from \textbf{(e)} to \textbf{(h)} with $\lambda_{\min}^{(m)} = \lambda^{(m)}$ indicated in the legends.
The detailed values of the scaling exponents are given in Tables~\ref{minimumenergyforM} and~\ref{maximumenergyforM}.
Other complementary cases are provided in Fig.~\ref{figsi_min1} to Fig.~\ref{figsi_max2}, and all parameters are the same as those in Fig.~\ref{energy_2max}.
 }
 \label{fig_Min_Max_high_NND_PD}
\end{figure}

Assuming for simplicity that the system starts at the origin ($\x_0 = \zero$), only the last snapshot matters because in principle, one can exploit the fact that until the final snapshot and from that point proceed to $\x_\text{f}$.
In this case, inner snapshots merely contribute to the exponent
$$\lambda_{\max} = \mathop {\max} \limits_{l} \left\{ \lambda_{\max}^{(M)} + \mathop {\sum_{m=l}^{M-1}} \limits_{1 \leq l \leq M-1}  \lambda_{\max}^{(m)} \right\}$$
that governs the exponential decrease of the energy for large $h$, where $\lambda_{\max}^{(m)}$ is the maximum eigenvalue of the snapshot $\A_m$.
For $\underline{E}$ (for $\overline{E}$, it is similar, and please see SI), when the last snapshot $\A_M$ is not negative definite ($\lambda_{\max}^{(M)} > 0$), $\underline{E}$ will decrease exponentially with an exponent between $\lambda_{\max}^{(M)}$ and $\lambda_{\max}$;
when $\A_M$ is negative definite ($\lambda_{\max}^{(M)} < 0$), $\underline{E}$ will decrease from a constant to exponentially with exponent $\lambda_{\max}$;
when $\A_M$ is negative semi-definite ($\lambda_{\max}^{(M)} = 0$),  $\underline{E}$ will decrease hyperbolically first and eventually exponentially with rate $\lambda_{\max}$.
Above analytical results are validated by numerical calculations (see Fig.~\ref{fig_Min_Max_high_NND_PD}, Fig.~\ref{figsi_min2} and Fig.~\ref{figsi_max2} in SI Sec.~\ref{EnergyforMnets}).
Finally, when $h$ is small, we predict that $\underline{E} \sim h^{-1}$, which is confirmed numerically by simulations and shown in Fig.~\ref{figsi_min1}. 
The detailed analytical scaling behavior of $\underline{E}$ and $\overline{E}$ for arbitrary number of snapshots and driver nodes may be found in SI Sec.~\ref{si_M_p}.

%Based on the minimal energy given in (\ref{{retemp}}), the upper bound $\overline{E}$ of $E^*$ is acquired with $\x_0 = \zero$ and $\| \x_\text{f} \| = 1$ theoretically.
%We find that, $E_\textrm{max} = 1/(2\lambda_\textrm{min})$, and it is actually the control energy $E^*$ for $\x_\text{f}$ along the direction of the eigenvector corresponded to the minimum eigenvalue $\lambda_\textrm{min}$ of $\s\W\s^{\textrm{T}}$.
%As the final state $\x_\text{f}$ is chosen randomly from the sphere centered on $\x_0$, it is the linear combination of all eigenvectors (containing the one of $\lambda_{\min}$) of $\s\W\s^{\textrm{T}}$.
%Thus the scaling of $\bar{E^*}$ is determined by $E_\textrm{max}$ (see Fig.~\ref{fig_energy}), which first decreases at $\bar{E^*} \sim \Delta t ^ {- \gamma}$ for small $\Delta t$ and then keeps as a constant (see SI ?).
%Furthermore, we could save control energy for both temporal and static networks by adding more driver nodes (see Fig.~\ref{fig_energy}).
%The energy difference between them decreases when we control all nodes directly (see Fig.~\ref{fig_energy}c), while temporal network still save energy when the duration time for each snapshot is large.

\section{Discussion}

Our results provide a comprehensive anatomy of the control energy scaling for undirected temporal networks with respect to the stability properties of the underlying system matrices.
Our results can readily be generalized to the case of weighted directed networks, provided the effective Gramian matrix is diagonalizable.
In this case, the traditional eigenvalues would be replaced by the real parts of the new (now complex) eigenvalues. 
%and the constant number in the undirected cases could be obtained by first order approximations of the Taylor expansions. 
In the present work each snapshot is confined to be controllable, 
%in order to compare the scaling behavior between temporal and static networks, since actually there is no certain relationship between the full controllability of them.
as it is difficult to perform a systematic equal comparison of the optimal control energy in temporal versus static networks if either of them is only partially controllable.
This is true in part because the optimal control trajectory may be highly nonlocal even as the distance between $\x_0$ and $\x_\text{f}$ approaches zero \cite{Sun2013prl,Li2016}.

The analysis of a single snapshot can provide intuition about why $\overline{E}$ and $\underline{E}$ are divided into three cases according to the properties of the final snapshot. 
For small $h$ (high temporality), the system has less overall time to allocate its optimal control scheme, meaning the last snapshot has correspondingly less influence over the scaling of both $\overline{E}$ and $\underline{E}$, thus explaining their broad power-law bahavior in this case.
For a final state chosen randomly from the controllable space, it has been shown the minimum control energy to reach it is dominated by the upper bound at the same control distance $\| \x_\text{f} - \x_0 \|$ for both temporal and static networks \cite{Yan2012PRL,Li2016}.
Our discovery of the scaling behavior of both $\overline{E}$ clearly explains the previous discovery \cite{Li2016} that temporal networks require orders of magnitude less control energy than their static counterparts, especially in the regime of high temporality (small $h$). 
Moreover, our analysis of $\underline{E}$ provide us the ``best case" control scenario at a given control distance.

To gain a deeper understanding of the scaling behavior of control energy for temporal networks, we can consider $\overline{E}$ as an example.
The optimal energy is inevitably affected by the internal system dynamics in the absence of control.
Indeed, for a single snapshot, the autonomous dynamics $\dot{\x}(t) = \A_m\x(t)$ will naturally facilitate movement away from the origin, when the system is unstable ($\A_m$ is PD, i.e.~$\lambda_{\min}^{(m)}>0$).
It follows that, when external control inputs corresponding to the maximal energy are applied,
%Thus, with external control inputs $\u(t)$, 
the control trajectory corresponding to the optimal maximum energy $\overline{E}$ will choose the least hindrance from the internal dynamics, namely the control direction along the eigenvector of $\lambda_{\min}^{(m)}$.
It is the facilitation of the internal dynamics that leads to the exponential decrease of $\overline{E}$ over large control time $h$.
When there exists at least one negative eigenvalue (say, $\lambda_{\min}^{(m)}<0$, meaning $\A_m$ is NPD), the optimal control path will take advantage of this and drag $\overline{E}$ to a larger value even though the system is unstable along other eigenvectors.
When $\lambda_{\min}^{(m)}=0$ ($\A_m$ is PSD), $\overline{E}$ will correspond to a trajectory aligned with the eigenvector of $\lambda_{\min}^{(m)}$ even with other positive eigenvalues, leading to the hyperbolic decay of $\overline{E}$.
$\underline{E}$ can be similarly understood for long snapshot durations (low temporality) by virtue of the attributes of  the spectral properties of the system matrix.
%NND, ND, and NSD of system matrix.

%For the temporal network with a series of snapshots, the overall system possesses the flexibility to achieve its optimal arrangement over every snapshot in principle.
Temporal networks are known to possess tremendous flexibility over static networks precisely because they allow exploitation of the most favorable dynamical features of many networks (snapshots) as opposed to just one.
Yet here, we have shown that the large-scale behavior of the control energy will be inevitably dominated by the final snapshot $\A_M$ during the last leg of the system's journey from $\x_{t_{M-1}}$ to $\x_{t_M} = \x_\text{f}$.
Thus, although it appears changing network structure is required for dramatic control advantages over static networks, the precise effects of temporality can nonetheless be understood by appealing to a \emph{single} snapshot.

\newpage
\renewcommand\arraystretch{1.2}
\begin{table}[H]
\caption{
Scaling behavior of $\underline{E}$ for a temporal network with two snapshots, $\A_1$ and $\A_2$, from $\x_0 = \zero$ to $\x_\text{f}$ with $p$ driver nodes.
$\lambda_{\max}^{(1)}$ and $\lambda_{\max}^{(2)}$ are the maximum eigenvalues of $\A_1$ and $\A_2$, respectively.
The scaling can be divided into three cases according to the sign of $\lambda_{\max}^{(2)}$, where
$\A_2$ is NND ($\lambda_{\max}^{(2)} > 0$),
$\A_2$ is ND ($\lambda_{\max}^{(2)} < 0$),
and $\A_2$ is NSD ($\lambda_{\max}^{(2)} = 0$).
$\mathcal{I} = \{i_1,i_2,\cdots,i_p \}$ is the set of $p$ driver nodes, and $\lambda_{\max} = \lambda_{\max}^{(1)} + \lambda_{\max}^{(2)}$.
The Heaviside step function $\textrm{H}(x)$ satisfies $\textrm{H}(x < 0) = 0$ and $\textrm{H}(x \geq 0) = 1$.
These analytical results are validated by numerical calculations, shown in Fig.~\ref{energy_2min}, and the corresponding panels are given as the last column.
The more general case of a temporal network with $M>2$ snapshots can be found in SI Sec.~\ref{EnergyforMnets}.
}
\center
\begin{tabular}{cccccc}
\hline \hline
 \multirow{2}*{$\A_2$} &  \multirow{2}*{small $h$}  &  \multicolumn{3}{c}{large $h$} & \multirow{1}*{Numerical}  \\  \cline{3-5}
 &  & $\lambda_{\max} < 0$   &   $\lambda_{\max} = 0$   &   $\lambda_{\max} > 0$ & results\\ \hline
NND & \multirow{3}*{$h^{-1}$} &\multicolumn{3}{c}{$\textrm{e}^{-2\left[\lambda_{\max}^{(1)}\textrm{H}\left(\lambda_{\max}^{(1)}\right) + \lambda_{\max}^{(2)}\right]h}$} & Fig.~\ref{energy_2min}a,~\ref{energy_2min}d\\
ND  &  & $-\frac{1}{\sum_{c \in I} \A_2^{-1}(c,c)}$ & $-\frac{1}{2-\sum_{c \in I}\A_2^{-1}(c,c)}$ & $\textrm{e}^{-2\lambda_{\max} h}$ & Fig.~\ref{energy_2min}b,~\ref{energy_2min}e\\
NSD &  & $h^{-1}$ & $(h+1)^{-1}$ &  $\textrm{e}^{-2\lambda_{\max}h}$ & Fig.~\ref{energy_2min}c,~\ref{energy_2min}f\\
\hline \hline
\end{tabular}
\label{Tab_min_2}
\end{table}

\renewcommand\arraystretch{1.2}
\begin{table}[H]
\caption{
Scaling behavior of $\overline{E}$ for a temporal network with two snapshots, $\A_1$ and $\A_2$, from $\x_0 = \zero$ to $\x_\text{f}$ with $p$ driver nodes.
$\lambda_{\min} = \lambda_{\min}^{(1)} + \lambda_{\min}^{(2)}$, and $\lambda_{\min}^{(1)}$ and $\lambda_{\min}^{(2)}$ are the minimum eigenvalues of $\A_1$ and $\A_2$, respectively.
The scaling can be divided into three cases according to the sign of $\lambda_{\min}^{(2)}$, where
$\A_2$ is PD ($\lambda_{\min}^{(2)} > 0$),
$\A_2$ is NPD ($\lambda_{\min}^{(2)} < 0$),
and $\A_2$ is PSD ($\lambda_{\min}^{(2)} = 0$).
Numerical calculations are also given in Fig.~\ref{energy_2max}.
The more general case of a temporal network with $M>2$ snapshots can be found in SI Sec.~\ref{EnergyforMnets}.
}
\center
\begin{tabular}{cccccc}
\hline \hline
 \multirow{2}*{$\A_2$} &  \multirow{2}*{small $h$}  &  \multicolumn{3}{c}{large $h$} & \multirow{1}*{Numerical}  \\  \cline{3-5}
 &  & $\lambda_{\min} < 0$   &   $\lambda_{\min} = 0$   &   $\lambda_{\min} > 0$ & results\\ \hline
PD & \multirow{3}*{$h^{-\gamma}$} &\multicolumn{3}{c}{$\textrm{e}^{-2\left[\lambda_{\min}^{(1)}\textrm{H}\left(\lambda_{\min}^{(1)}\right) + \lambda_{\min}^{(2)}\right]h}$} & Fig.~\ref{energy_2max}a,~\ref{energy_2max}d\\
NPD  &  & $\sum_{c \in I} C(\A_2,c)$ & $\sum_{c \in I} C(\A_2,c)$ & $\textrm{e}^{-2\lambda_{\min} h}$ & Fig.~\ref{energy_2max}b,~\ref{energy_2max}e\\
PSD &  & $h^{-\gamma}$ & $(h+1)^{-1}$ &  $\textrm{e}^{-2\lambda_{\min}h}$ & Fig.~\ref{energy_2max}c,~\ref{energy_2max}f\\
\hline \hline
\end{tabular}
\label{Tab_max_2}
\end{table}

\newpage

\newpage
\setcounter{figure}{0}
\setcounter{table}{0}
\renewcommand{\thefigure}{S\arabic{figure}}
\renewcommand{\thetable}{S\arabic{table}}
\section{Supplementary Information}
\appendix

\section{Control energy for two snapshots and one driver node from $\x _0 = \zero$ to $\x_\text{f}$}
\label{Energyfor2netsReach}

We denote by $\A (i,j) = a_{ij}$ the entry at $i$th row and $j$th column in a matrix $\A $,
and let $\A _1=\left(a_{ij}\right)_{N \times N}$ and $\A _2=\left(b_{ij}\right)_{N \times N}$, where $N \times N$ represents the size of the corresponding matrix.
We assume without loss of generality that it is the $c$-th node that receives direct input, meaning we have $\B ^{\textrm{T}} = (0, \cdots, 1,\cdots,0)$ where the $c$th entry is $1$ while others are $0$.

When the two snapshots of the temporal network are undirected, the corresponding dynamical matrices $\A _1$ and $\A _2$ are symmetric, 
%For an undirected temporal network with two snapshots, and $\A _1$ and $\A _2$ are symmetric matrix with real numbers, 
allowing us to write $\A _1 = \textrm{\textbf{P}} \mathbf{\Theta} \textrm{\textbf{P}}^\textrm{T}$ and $\A _2 = \textrm{\textbf{Q}} \mathbf{\Gamma} \textrm{\textbf{Q}}^\textrm{T}$, where $\textrm{\textbf{P}}=(P_{ij})_{N \times N}$, $\textrm{\textbf{Q}}=(Q_{ij})_{N \times N}$,
$\mathbf{\Theta} = \text{diag}(\theta_1, \theta_2, \cdots, \theta_N)$, and $\mathbf{\Gamma} = \text{diag}(\gamma_1,\gamma_2, \cdots, \gamma_N)$.
%$\mathbf{\Theta} = \left(
%  \begin{array}{ccc}
%\theta_1 & & \\
%& \ddots & \\
%& & \theta_N
%  \end{array}
%\right)$,
%and
%$\mathbf{\Gamma} = \left(
%  \begin{array}{ccc}
%\gamma_1 & & \\
%& \ddots & \\
%& & \gamma_N
%  \end{array}
%\right)$.
Here $\theta_i$ ($\gamma_i$) are the (real) eigenvalues of $\A_1$ ($\A_2$),
and we assume $\theta_1 \geq \theta_2\geq\cdots\geq\theta_N$, $\gamma_1 \geq \gamma_2\geq\cdots\geq\gamma_N$.

As we control the temporal network from $\x_0=\zero$ to $\x_\text{f}$, we have that the effective gramian matrix is
\begin{eqnarray*}
\s\W\s^\textrm{T}  &=&
\underbrace{\textrm{e}^{\A _{2}h}
\cdot
\int_{0}^{h}\textrm{e}^{\A _{1}t}\B\B^\textrm{T}
\textrm{e}^{\A ^\textrm{T}_{1}t}\mathrm{d}t
\cdot
\textrm{e}^{\A ^\textrm{T}_{2}h}}_{\textrm{\textbf{R}}_1}
+
\underbrace{\int_{0}^{h}\textrm{e}^{\A _{2}t}\B\B^\textrm{T}
\textrm{e}^{\A ^\textrm{T}_{2}t}\mathrm{d}t}_{\textrm{\textbf{R}}_2},
\end{eqnarray*}
for which we can expand the two component terms using the above eigendecompositions as
\begin{eqnarray*}
\textrm{\textbf{R}}_1  &=&
\textrm{\textbf{Q}} \textrm{e}^{\mathbf{\Gamma} h} \textrm{\textbf{Q}} ^\textrm{T}
\textrm{\textbf{P}} \int_{0}^{h}\textrm{e}^{\mathbf{\Theta} t} \textrm{\textbf{P}}^\textrm{T}  \B \B^\textrm{T} \textrm{\textbf{P}}
\textrm{e}^{\mathbf{\Theta} t}\mathrm{d}t \textrm{\textbf{P}}^\textrm{T}
\textrm{\textbf{Q}} \textrm{e}^{\mathbf{\Gamma} h} \textrm{\textbf{Q}}^\textrm{T},
 \\
\textrm{\textbf{R}}_2
 &=&
\textrm{\textbf{Q}} \int_{0}^{h}\textrm{e}^{\mathbf{\Gamma} t} \textrm{\textbf{Q}}^\textrm{T} \B\B^\textrm{T} \textrm{\textbf{Q}}
\textrm{e}^{\mathbf{\Gamma} t}\mathrm{d}t \textrm{\textbf{Q}}^\textrm{T},
\end{eqnarray*}
%with
%\begin{eqnarray}
%\textrm{e}^{\mathbf{\Theta} t} P^\textrm{T} \B\B^\textrm{T} P   \textrm{e}^{\mathbf{\Theta} t}
%& = &
%\left(
%  \begin{array}{ccc}
%\textrm{e}^{\theta_1 t} & & \\
%& \ddots & \\
%& & \textrm{e}^{\theta_N t}
%  \end{array}
%\right)
%\left(
%  \begin{array}{c}
%P_{c1} \\
%\vdots \\
%P_{cN}
%  \end{array}
%\right)
%\left(
%  \begin{array}{ccc}
%P_{c1}& \cdots & P_{cN}
%  \end{array}
%\right)
%\left(
%  \begin{array}{ccc}
%\textrm{e}^{\theta_1 t} & & \\
%& \ddots & \\
%& & \textrm{e}^{\theta_N t}
%  \end{array}
%\right)
%\nonumber \\
%& = &
%\left(  \textrm{e}^{(\theta_i+\theta_j) t} P_{ci}P_{cj} \right)_{N \times N}
%\end{eqnarray}
%and then we obtain
which results in
\begin{eqnarray}
\label{R1ij}
\textrm{\textbf{R}}_1(i,j)
 &=&
\sum_{r=1}^{N} \sum_{s=1}^{N} \sum_{m=1}^{N} \sum_{n=1}^{N}
Q_{ir} \textrm{e}^{\gamma_r h} Q_{sr}
\left\{
\sum_{k=1}^{N}\sum_{l=1}^{N} \frac{P_{sk}P_{ck}P_{cl}P_{ml}}{\theta_k+\theta_l}\left[ \textrm{e}^{(\theta_k+\theta_l)h}-1\right]
\right\}
Q_{mn} \textrm{e}^{\gamma_n h} Q_{jn}
\nonumber \\
& = &
\sum_{r=1}^{N} \sum_{s=1}^{N} \sum_{m=1}^{N} \sum_{n=1}^{N}
Q_{ir} \textrm{e}^{( \gamma_r + \gamma_n ) h} Q_{sr} Q_{mn}  Q_{jn}
\sum_{k=1}^{N}\sum_{l=1}^{N} \frac{P_{sk}P_{ck}P_{cl}P_{ml}}{\theta_k+\theta_l}\left[ \textrm{e}^{(\theta_k+\theta_l)h}-1\right],
\\
\textrm{\textbf{R}}_2(i,j)
 &=&
\sum_{k=1}^{N}\sum_{l=1}^{N} \frac{Q_{ik}Q_{ck}Q_{cl}Q_{jl}}{\gamma_k+\gamma_l}\left[ \textrm{e}^{(\gamma_k+\gamma_l)h}-1\right].
\label{R2ij}
\end{eqnarray}
This allows us to analyze $\textrm{\textbf{R}}_1$ and $\textrm{\textbf{R}}_2$ in terms of the magnitude of $h$ as follows:

\subsection{As $h \rightarrow 0$ }
When $h \rightarrow 0$, we can make the approximation $\textrm{e}^{(\gamma_k+\gamma_l)h} \approx 1 + (\gamma_k+\gamma_l)h$.
Then we have
\begin{eqnarray}
\textrm{\textbf{R}}_2(i,j)
 & \approx &
\sum_{k=1}^{N}\sum_{l=1}^{N} Q_{ik}Q_{ck}Q_{cl}Q_{jl} h
\nonumber \\
& = &
\begin{cases}
h\sum_{k=1}^{N} Q_{ik}Q_{ck}
      & \text{if $j=c$}  \\
h\sum_{l=1}^{N}Q_{cl}Q_{jl}
      & \text{if $i=c$}    \\
0
      & \text{otherwise}      
 \end{cases}
\nonumber \\
& = &
\begin{cases}
h
      & \text{if $i=j=c$}  \\
0
      & \text{otherwise}
 \end{cases}
\label{R2ijsh}
\\
\textrm{\textbf{R}}_1(i,j)
 & \approx &
\sum_{r=1}^{N} \sum_{s=1}^{N} \sum_{m=1}^{N} \sum_{n=1}^{N}
Q_{ir} \textrm{e}^{( \gamma_r + \gamma_n ) h} Q_{sr} Q_{mn}  Q_{jn}
h   ~~~~~~~~  \text{(here $s=m=c$)}
\nonumber  \\
& = &
\sum_{r=1}^{N}\sum_{n=1}^{N}
Q_{ir}  Q_{cr} Q_{cn}  Q_{jn}
\left[ 1+ ( \gamma_r + \gamma_n ) h \right] h
\label{R1ijsh} \\
& = &
\underbrace{h\sum_{r=1}^{N}\sum_{n=1}^{N} Q_{ir}  Q_{cr} Q_{cn}  Q_{jn}}_{\Omega_1}
+
\underbrace{h^2  \sum_{r=1}^{N}\sum_{n=1}^{N} Q_{ir}  Q_{cr} Q_{cn}  Q_{jn} \gamma_r } _{\Omega_2}
+
\underbrace{h^2  \sum_{r=1}^{N}\sum_{n=1}^{N} Q_{ir}  Q_{cr} Q_{cn}  Q_{jn} \gamma_n} _{\Omega_3},
\nonumber
\end{eqnarray} 
where the three terms in the final expression obey
\begin{eqnarray*}
\Omega_1  & = &
 \begin{cases}
h
      & \text{if $i=j=c$}  \\
0
      & \text{otherwise}
 \end{cases}, ~~~~~
\Omega_2   = 
 \begin{cases}
h^2  \sum_{r=1}^{N} Q_{ir}  Q_{cr}  \gamma_r
= h^2 b_{ic}
      & \text{if $j=c$}  \\
0
      & \text{otherwise}
 \end{cases},
 \\
\Omega_3 & = &
 \begin{cases}
h^2  \sum_{n=1}^{N} Q_{cn}  Q_{jn}  \gamma_n
= h^2 b_{cj}
      & \text{if $i=c$}  \\
0
      & \text{otherwise}
 \end{cases}.
\end{eqnarray*}

Thus we have
\begin{eqnarray*}
\textrm{\textbf{R}}_1(i,j) + \textrm{\textbf{R}}_2(i,j) =  
 \begin{cases}
2h+ 2b_{cc}h^2
      & \text{if $i=j=c$}  \\
b_{cj}h^2
      & \text{if $i=c$ and $j \neq c$} \\
b_{ic}h^2
      & \text{if $j=c$ and $i \neq c$}
 \end{cases},
\end{eqnarray*}
and by adding $\textrm{\textbf{R}}_1$ and $\textrm{\textbf{R}}_2$ we obtain  
\begin{eqnarray*}
\s\W\s^\textrm{T}  =
 \left(
  \begin{array}{ccccccc}
0 & \cdots & 0 & b_{1c}h^2 & 0 & \cdots & 0  \\
\vdots & &\vdots & \vdots &\vdots & & \vdots \\
0 & \cdots & 0 & b_{i-1,c}h^2 & 0 & \cdots & 0  \\
b_{c1}h^2 & \cdots & b_{c,i-1}h^2 & 2h+2b_{cc}h^2 & b_{c,i+1}h^2 & \cdots & b_{cN}h^2  \\
0 & \cdots & 0 & b_{i+1,c}h^2 & 0 & \cdots & 0  \\
\vdots & &\vdots & \vdots& \vdots & & \vdots \\
0 & \cdots & 0 & b_{Nc}h^2 & 0 & \cdots & 0  \\
  \end{array}
\right).
\end{eqnarray*}
As for the associated eigenvalues, we must solve the following equations
\begin{eqnarray}
|\s\W\s^\textrm{T} - \lambda \textrm{\textbf{I}}|  =
 \left |
  \begin{array}{ccccccc}
-\lambda & \cdots & 0 & b_{1c}h^2 & 0 & \cdots & 0  \\
\vdots & &\vdots & \vdots &\vdots & & \vdots \\
0 & \cdots & -\lambda & b_{i-1,c}h^2 & 0 & \cdots & 0  \\
b_{c1}h^2 & \cdots & b_{c,i-1}h^2 & 2h+2b_{cc}h^2 - \lambda& b_{c,i+1}h^2 & \cdots & b_{cN}h^2  \\
0 & \cdots & 0 & b_{i+1,c}h^2 & -\lambda & \cdots & 0  \\
\vdots & &\vdots & \vdots& \vdots & & \vdots \\
0 & \cdots & 0 & b_{Nc}h^2 & 0 & \cdots & -\lambda \\
  \end{array}
\right |
 ~~~~~~~~~~~~~~~~~~~~
\nonumber \\
 = 
 \left |
  \begin{array}{ccccccc}
-\lambda & \cdots & 0 & 0 & 0 & \cdots & 0  \\
\vdots & &\vdots & \vdots &\vdots & & \vdots \\
0 & \cdots & -\lambda & 0 & 0 & \cdots & 0  \\
b_{c1}h^2 & \cdots & b_{c,i-1}h^2 &
2h+2b_{cc}h^2 - \lambda   + \frac{h^4}{\lambda}\sum_{i=1,i \neq c}^{N} b_{ic}b_{ci}
 & b_{c,i+1}h^2 & \cdots & b_{cN}h^2  \\
0 & \cdots & 0 & 0 & -\lambda & \cdots & 0  \\
\vdots & &\vdots & \vdots& \vdots & & \vdots \\
0 & \cdots & 0 & 0 & 0 & \cdots & -\lambda \\
  \end{array}
\right |
 ~~~~~~~~~~
%\nonumber \\
%& = &
%\left( 2h+2b_{cc}h^2 - \lambda  + \frac{h^4}{\lambda}\sum_{i=1,i \neq c}^{N} b_{ic}^2 \right) (-\lambda)^{N-1}
\nonumber \\
 = 
\left[ \lambda^2- \left( 2h+2b_{cc}h^2 \right)\lambda  - h^4\sum_{i=1,i \neq c}^{N} b_{ic}^2 \right] (-\lambda)^{N-2}
 = 0.
 ~~~~~~~~~~~~~~~~~~~~
  ~~~~~~~~~~~~~~~~~~~~
   ~~~~~~~~~~~
\end{eqnarray}
This yields the approximated eigenvalues as $\lambda=0$ (with multiplicity $N-2$), $\lambda=h+b_{cc}h^2 \pm \sqrt{\left( h+b_{cc}h^2 \right)^2 + h^4\sum_{i=1,i \neq c}^{N} b_{ic}^2 }$, and thus
\begin{equation*}
\lambda_{\max}=h+b_{cc}h^2 + \sqrt{h^2+2b_{cc}h^3  + h^4\sum_{i=1}^{N} b_{ic}^2 }.
\end{equation*}
Therefore, in this case, i.e., $h \rightarrow 0$, we have
\begin{equation*}
\Elb \approx \frac{1}{2\left( h+b_{cc}h^2 + \sqrt{h^2+2b_{cc}h^3  + h^4\sum_{i=1}^{N} b_{ic}^2 }\right)}.
\end{equation*}

\subsection{For large $h$}
For a square matrix, the trace of the matrix is the sum of the eigenvalues.
Here when $h$ is large, we use the trace of $\s\W\s^\textrm{T} $ to approximate its maximum eigenvalue, i.e.,
\begin{equation*}
\lambda_{\max} \approx \textrm{Tr}(\s\W\s^\textrm{T}) = \sum_{i=1}^{N} \left( \textrm{\textbf{R}}_1(i,i) + \textrm{\textbf{R}}_2(i,i)  \right).
\end{equation*}
We have
\begin{eqnarray*}
\sum_{i=1}^{N} \textrm{\textbf{R}}_2(i,i)
 & = &
\sum_{k=1}^{N}\sum_{l=1}^{N} \frac{Q_{ck}Q_{cl}}{\gamma_k+\gamma_l}\left[ \textrm{e}^{(\gamma_k+\gamma_l)h}-1\right]
\sum_{i=1}^{N}Q_{ik}Q_{il}
\nonumber \\
& = &
\sum_{k=1}^{N} \frac{Q_{ck}^2 }{2 \gamma_k}\left[ \textrm{e}^{2 \gamma_k h}-1\right]
~~~~~~~~~~\left( \text{note that if $k \neq l,~~~\sum_{i=1}^{N}Q_{ik}Q_{il} = 0$ } \right)
\nonumber \\
& \approx &
\begin{cases}
- \frac{1}{2}\sum_{k=1}^{N} Q_{ck}^2 \gamma_k ^ {-1}=  - \frac{1}{2} A_2^{-1}(c,c)
& \text{if $\A _2$ is Negative Definite (ND)}  \\
h\sum_{k=1}^{N} Q_{ck}^2 = h
& \text{if $\A _2$ is Negative Semi Definite (NSD)}   \\
\textrm{e}^{2 \gamma_1 h}
& \text{otherwise, i.e., if $\A _2$ is Not Negative Definite (NND)}
 \end{cases}
\end{eqnarray*}
\begin{eqnarray*}
\sum_{i=1}^{N} \textrm{\textbf{R}}_1(i,i)
 & = &
\sum_{r=1}^{N} \sum_{s=1}^{N} \sum_{m=1}^{N} \sum_{n=1}^{N}
 \textrm{e}^{( \gamma_r + \gamma_n ) h} Q_{sr} Q_{mn}
\sum_{k=1}^{N}\sum_{l=1}^{N} \frac{P_{sk}P_{ck}P_{cl}P_{ml}}{\theta_k+\theta_l}\left[ \textrm{e}^{(\theta_k+\theta_l)h}-1\right]
\sum_{i=1}^{N} Q_{ir}Q_{in}
\nonumber \\
& = &
\sum_{r=1}^{N} \sum_{s=1}^{N} \sum_{m=1}^{N}
 \textrm{e}^{2 \gamma_r  h} Q_{sr} Q_{mr}
\sum_{k=1}^{N}\sum_{l=1}^{N} \frac{P_{sk}P_{ck}P_{cl}P_{ml}}{\theta_k+\theta_l}\left[ \textrm{e}^{(\theta_k+\theta_l)h}-1\right]
  \left( \text{, if $r \neq n,~~~\sum_{i=1}^{N} Q_{ir}Q_{in} = 0$ } \right)
%\nonumber \\
\end{eqnarray*}
\begin{eqnarray*}
& = &
\sum_{s=1}^{N} \sum_{m=1}^{N} W_1(s,m)
\textrm{e}^{2A_2h}(s,m)
\nonumber \\ &  &  
 \Big( \text{note that $\sum_{r=1}^{N} \textrm{e}^{2 \gamma_r  h} Q_{sr} Q_{mr} = \textrm{e}^{2A_2h}(s,m)$}
\nonumber \\ &  & 
\text{and $\sum_{k=1}^{N}\sum_{l=1}^{N} \frac{P_{sk}P_{ck}P_{cl}P_{ml}}{\theta_k+\theta_l}\left[ \textrm{e}^{(\theta_k+\theta_l)h}-1\right] = W_1(s,m) $} \Big)
\nonumber \\
& \approx &
\begin{cases}
\sum_{r=1}^{N} \sum_{s=1}^{N} \sum_{m=1}^{N}
 \textrm{e}^{2 \gamma_r  h} Q_{sr} Q_{mr}
& \text{if $\A _1$ is ND or NSD} \\
\textrm{e}^{2 \gamma_1 h} \textrm{e}^{2 \theta_1h}
& \text{if $\A _1$ is NND}
 \end{cases}
\end{eqnarray*}

Hence we obtain that
%former form
%\begin{eqnarray}
%\lambda_{\max}
%& \approx &
%\begin{cases}
% - \frac{1}{2} A_2^{-1}(c,c)
% & \text{if $\A_1$ is ND and $\A_2$ is ND}  \\
% h
% & \text{if $\A_1$ is ND and $\A_2$ is NSD}  \\
%2\textrm{e}^{2 \gamma_1 h}
% & \text{if $\A_1$ is ND and $\A_2$ is NND}  \\
%  - \frac{1}{2} A_2^{-1}(c,c)
% & \text{if $\A_1$ is NSD and $\A_2$ is ND}  \\
% h
% & \text{if $\A_1$ is NSD and $\A_2$ is NSD}  \\
%2 \textrm{e}^{2 \gamma_1 h}
% & \text{if $\A_1$ is NSD and $\A_2$ is NND}  \\
%\textrm{e}^{2 \gamma_1 h} \textrm{e}^{(\theta_1+\theta_2)h}  - \frac{1}{2} A_2^{-1}(c,c)
% & \text{if $\A_1$ is NND and $\A_2$ is ND}  \\
%\textrm{e}^{2 \gamma_1 h} \textrm{e}^{(\theta_1+\theta_2)h} +h
% & \text{if $\A_1$ is NND and $\A_2$ is NSD}  \\
%\textrm{e}^{2 \gamma_1 h} \textrm{e}^{(\theta_1+\theta_2)h} + \textrm{e}^{2 \gamma_1 h}
% & \text{if $\A_1$ is NND and $\A_2$ is NND}
% \end{cases}
%\end{eqnarray}
\begin{eqnarray*}
\lambda_{\max}
& \approx &
\begin{cases}
2 \textrm{e}^{2 \gamma_1 h}
 & \text{if $\A _1$ is ND or NSD and $\A _2$ is NND}  \\
\textrm{e}^{2 \gamma_1 h} \textrm{e}^{2 \theta_1 h} + \textrm{e}^{2 \gamma_1 h}
 & \text{if $\A _1$ is NND and $\A _2$ is NND} \\
 h
 & \text{if $\A _1$ is ND or NSD and $\A _2$ is NSD}  \\
\textrm{e}^{2 \gamma_1 h} \textrm{e}^{2 \theta_1 h} +h
 & \text{if $\A _1$ is NND and $\A _2$ is NSD}  \\
  - \frac{1}{2} \A_2^{-1}(c,c)
 & \text{if $\A _1$ is ND or NSD and $\A _2$ is ND}  \\
\textrm{e}^{2 \gamma_1 h} \textrm{e}^{2 \theta_1 h}  - \frac{1}{2} \A_2^{-1}(c,c)
 & \text{if $\A _1$ is NND and $\A _2$ is ND}
 \end{cases}.
\end{eqnarray*}

Therefore, the scaling of $\Elb$  for controlling temporal networks from $\x _0 = \textrm{\textbf{0}}$ to $\x_\text{f}$ is
\begin{eqnarray*}
\Elb \sim
\begin{cases}
 h^{-1}
 ~~~ \text{small $h$}  \\
 \frac{\text{large $h$, $\A_2$ is NND}}{\text{decreasing exponentially}}
  \begin{cases}
   \textrm{e}^{-2 \gamma_1 h}
 & \text{if $\theta_1 \leq 0$  }  \\
  \textrm{e}^{- 2(\gamma_1+\theta_1) h}
 & \text{if $\theta_1 > 0$  }
 \end{cases} \\
 \frac{\text{large $h$, $\A_2$ is ND}}{\text{decreasing from constant to exponentially}}
  \begin{cases}
    - \frac{1}{\A_2^{-1}(c,c)}
 & \text{if $\gamma_1+\theta_1< 0$}  \\
     \frac{1}{2 - \A_2^{-1}(c,c)}
 & \text{if $\gamma_1+\theta_1 =0$}  \\
  \textrm{e}^{- 2(\gamma_1+\theta_1) h}
 & \text{if $\gamma_1+\theta_1 > 0$}
 \end{cases} \\
  \frac{\text{large $h$, $\A_2$ is NSD}}{\text{decreasing from hyperbolically to exponentially}}
  \begin{cases}
 h^{-1}
~~~ \text{if $\theta_1 \leq 0$ }  \\
 \textrm{e}^{-2\theta_1h}
 ~~~ \text{if $\theta_1  > 0$ }
  \end{cases}
 \end{cases}.
\end{eqnarray*}

From the numerical calculations, we have the scaling of $\Eub$  for controlling temporal networks from $\x _0 = \textrm{\textbf{0}}$ to $\x_\text{f}$ is
\begin{eqnarray*}
\Eub \sim
\begin{cases}
 h^{-N}
 ~~~ \text{small $h$}  \\
 \frac{\text{large $h$, $\A_2$ is PD}}{\text{decreasing exponentially}}
  \begin{cases}
   \textrm{e}^{-2 \gamma_N h}
 & \text{if $\theta_N \leq 0$  }  \\
  \textrm{e}^{- 2(\gamma_N+\theta_N) h}
 & \text{if $\theta_N > 0$  }
 \end{cases} \\
 \frac{\text{large $h$, $\A_2$ is NPD}}{\text{decreasing from constant to exponentially}}
  \begin{cases}
     C(\A_2, c)
 & \text{if $\gamma_N+\theta_N \leq 0$}  \\
  \textrm{e}^{- 2(\gamma_N+\theta_N) h}
 & \text{if $\gamma_N+\theta_N > 0$}
 \end{cases} \\
  \frac{\text{large $h$, $\A_2$ is PSD}}{\text{decreasing from hyperbolically to exponentially}}
  \begin{cases}
 h^{-N}
~~~ \text{if $\theta_N \leq 0$ }  \\
 \textrm{e}^{-2\theta_Nh}
 ~~~ \text{if $\theta_N  > 0$ }
  \end{cases}
 \end{cases}.
\end{eqnarray*}

%\newpage
\section{Control energy for two snapshots and one driver node from $\x _0$ to $\x_\text{f} = \textrm{\textbf{0}}$}
\label{Energyfor2netsControl}
%
%As $\x_\textbf{\textrm{0}} \neq 0$ and $\x_\text{f} = \textbf{\textrm{0}}$ (which is called controllability in control theory), the normalized minimum energy is
%\begin{eqnarray}
%E^*(\x _0, \zero)
%& = & \frac{\x_0^\textrm{T}  \left( \textbf{\textrm{S}}\W\textbf{\textrm{S}}^\textrm{T} \right) ^ {-1} \x_0} {2\x_0^\textrm{T}\x_0}
%\end{eqnarray}
%where $\textbf{\textrm{S}}= \left(  \mathrm{e}^{-\A _{1}h_1} , \cdots,  \prod_{l=1}^{j} \mathrm{e}^{-\A _{l}h_l}  ,\cdots,  \prod_{l=1}^M \mathrm{e}^{-\A _{l}h_l} \right)$.
%Still we have
%\begin{eqnarray}
%\frac{1}{2\lambda^*_{\max}} = \Elb \leq  E^*(\x _0, \zero) \leq  \Eub = \frac{1}{2\lambda^*_{\min}}
%\end{eqnarray}
%for every $\x _0$, where $\lambda^*_{\max}$ and $\lambda^*_{\min}$  are the maximum and minimum eigenvalue of $\textbf{\textrm{S}}\W\textbf{\textrm{S}}^\textrm{T}$.

If there are two snapshots $\A _1$ and $\A _2$, and $\x_\text{f}=\zero$, we can follow a similar procedure to the above to write
\begin{eqnarray*}
\s\W\s^\textrm{T} 
 = 
\underbrace{\textrm{e}^{-\A _{1}h}
\cdot
\int_{0}^{h}\textrm{e}^{\A _{1}t}\B\B^\textrm{T}
\textrm{e}^{\A ^\textrm{T}_{1}t}\mathrm{d}t
\cdot
\textrm{e}^{-\A ^\textrm{T}_{1}h}}_{\textbf{\textrm{C}}_1}
+
\underbrace{\textrm{e}^{-\A _{1}h}\textrm{e}^{-\A _{2}h}
\cdot
\int_{0}^{h}\textrm{e}^{\A _{2}t}\B\B^\textrm{T}
\textrm{e}^{\A ^\textrm{T}_{2}t}\mathrm{d}t
\cdot
\textrm{e}^{-\A ^\textrm{T}_{2}h}
\textrm{e}^{-\A ^\textrm{T}_{1}h}}_{\textbf{\textrm{C}}_2},
\nonumber
\end{eqnarray*}
where the individual terms can be expanded as
\begin{eqnarray*}
\textbf{\textrm{C}}_1  &=&
\textbf{\textrm{P}} \textrm{e}^{- \mathbf{\Theta} h} \textbf{\textrm{P}}^\textrm{T}
\textbf{\textrm{P}} \int_{0}^{h}\textrm{e}^{\mathbf{\Theta} t} \textbf{\textrm{P}}^\textrm{T}  \B \B^\textrm{T}
\textbf{\textrm{P}} \textrm{e}^{\mathbf{\Theta} t}\mathrm{d}t \textbf{\textrm{P}}^\textrm{T}
\textbf{\textrm{P}} \textrm{e}^{- \mathbf{\Theta} h} \textbf{\textrm{P}}^\textrm{T}
=
\textbf{\textrm{P}} \textrm{e}^{- \mathbf{\Theta} h}
 \int_{0}^{h}\textrm{e}^{\mathbf{\Theta} t} \textbf{\textrm{P}}^\textrm{T}  \B \B^\textrm{T}
\textbf{\textrm{P}} \textrm{e}^{\mathbf{\Theta} t}\mathrm{d}t
 \textrm{e}^{- \mathbf{\Theta} h} \textbf{\textrm{P}}^\textrm{T},
 \\
\textbf{\textrm{C}}_2
 &=&
\textbf{\textrm{P}} \textrm{e}^{- \mathbf{\Theta} h} \textbf{\textrm{P}}^\textrm{T}
\textrm{e}^{-\A _{2}h}
\int_{0}^{h}\textrm{e}^{\A _{2}t}\B\B^\textrm{T}
\textrm{e}^{\A ^\textrm{T}_{2}t}\mathrm{d}t
\textrm{e}^{-\A ^\textrm{T}_{2}h}
\textbf{\textrm{P}} \textrm{e}^{- \mathbf{\Theta} h} \textbf{\textrm{P}}^\textrm{T}.
\end{eqnarray*}
From the following relation
\begin{eqnarray*}
\textrm{e}^{\mathbf{\Theta} t} \textbf{\textrm{P}}^\textrm{T} \B\B^\textrm{T} \textbf{\textrm{P}}   \textrm{e}^{\mathbf{\Theta} t}
& = &
\left(
  \begin{array}{ccc}
\textrm{e}^{\theta_1 t} & & \\
& \ddots & \\
& & \textrm{e}^{\theta_N t}
  \end{array}
\right)
\left(
  \begin{array}{c}
P_{c1} \\
\vdots \\
P_{cN}
  \end{array}
\right)
\left(
  \begin{array}{ccc}
P_{c1}& \cdots & P_{cN}
  \end{array}
\right)
\left(
  \begin{array}{ccc}
\textrm{e}^{\theta_1 t} & & \\
& \ddots & \\
& & \textrm{e}^{\theta_N t}
  \end{array}
\right)
\nonumber \\
& = &
\left(  \textrm{e}^{(\theta_i+\theta_j) t} P_{ci}P_{cj} \right)_{N \times N},
\end{eqnarray*}
we have
\begin{eqnarray*}
\textrm{\textbf{C}}_1(i,j)
 &=&
\sum_{k=1}^{N}\sum_{l=1}^{N} P_{ik}  \textrm{e}^{- \theta_k h}
 \frac{P_{ck}P_{cl}}{\theta_k+\theta_l}\left[\textrm{e}^{(\theta_k+\theta_l)h} - 1\right]
  \textrm{e}^{- \theta_l h}   P_{jl}
\nonumber \\
& = &
\sum_{k=1}^{N}\sum_{l=1}^{N} \frac{P_{ik}P_{ck}P_{cl}P_{jl}}{\theta_k+\theta_l}\left[1- \textrm{e}^{- (\theta_k+\theta_l)h}\right],
\\
\textrm{\textbf{C}}_2(i,j)
 &=&
\sum_{r=1}^{N} \sum_{s=1}^{N} \sum_{m=1}^{N} \sum_{n=1}^{N}
P_{ir} \textrm{e}^{-\theta_r h} P_{sr}
\left\{
\sum_{k=1}^{N}\sum_{l=1}^{N} \frac{Q_{sk}Q_{ck}Q_{cl}Q_{ml}}{\gamma_k+\gamma_l}\left[ 1- \textrm{e}^{-(\gamma_k+\gamma_l)h}\right]
\right\}
P_{mn} \textrm{e}^{-\theta_n h} P_{jn}
\nonumber \\
& = &
\sum_{r=1}^{N} \sum_{s=1}^{N} \sum_{m=1}^{N} \sum_{n=1}^{N}
P_{ir} \textrm{e}^{- ( \theta_r + \theta_n ) h} P_{sr} P_{mn}  P_{jn}
\sum_{k=1}^{N}\sum_{l=1}^{N} \frac{Q_{sk}Q_{ck}Q_{cl}Q_{ml}}{\gamma_k+\gamma_l}\left[ 1- \textrm{e}^{-(\gamma_k+\gamma_l)h}\right].
\end{eqnarray*}

This allows us to analyze $\textbf{\textrm{C}}_1$ and $\textbf{\textrm{C}}_2$ according to the magnitude of $h$.

\subsection{As $h \rightarrow 0$ }
By making the approximation $\textrm{e}^{-(\gamma_k+\gamma_l)h} \approx 1 - (\gamma_k+\gamma_l)h$,
we have
\begin{eqnarray*}
\textrm{\textbf{C}}_1(i,j)
 & \approx &
\sum_{k=1}^{N}\sum_{l=1}^{N} P_{ik}P_{ck}P_{cl}P_{jl} h
\nonumber \\
& = &
\begin{cases}
h\sum_{k=1}^{N} P_{ik}P_{ck}
      & \text{if $j=c$}  \\
h\sum_{l=1}^{N}P_{cl}P_{jl}
      & \text{if $i=c$} \\
0
      & \text{otherwise}
 \end{cases}
%\nonumber \\
= 
\begin{cases}
h
      & \text{if $i=j=c$}  \\
0
      & \text{otherwise}
 \end{cases},
\\
\textrm{\textbf{C}}_2(i,j)
 & \approx &
\sum_{r=1}^{N} \sum_{s=1}^{N} \sum_{m=1}^{N} \sum_{n=1}^{N}
P_{ir} \textrm{e}^{-( \theta_r + \theta_n ) h} P_{sr} P_{mn}  P_{jn}
h   ~~~~~~~~  (here ~~s=m=c)
 \\
& = &
\sum_{r=1}^{N}\sum_{n=1}^{N}
P_{ir}  P_{cr} P_{cn}  P_{jn}
\left[ 1- ( \theta_r + \theta_n ) h \right] h
\nonumber \\
& = &
\underbrace{h\sum_{r=1}^{N}\sum_{n=1}^{N} P_{ir}  P_{cr} P_{cn}  P_{jn}}_{\Delta_1}
-
\underbrace{h^2  \sum_{r=1}^{N}\sum_{n=1}^{N} P_{ir}  P_{cr} P_{cn}  P_{jn} \theta_r } _{\Delta_2}
-
\underbrace{h^2  \sum_{r=1}^{N}\sum_{n=1}^{N} P_{ir}  P_{cr} P_{cn}  P_{jn} \theta_n} _{\Delta_3}.
\nonumber
\end{eqnarray*}
 Furthermore, we obtain
\begin{eqnarray*}
\Delta_1  & = &
 \begin{cases}
h
      & \text{if $i=j=c$}  \\
0
      & \text{otherwise}
 \end{cases},
 \end{eqnarray*}
 \begin{eqnarray*}
\Delta_2  & = &
 \begin{cases}
-h^2  \sum_{r=1}^{N} P_{ir}  P_{cr}  \theta_r
= -h^2 a_{ic}
      & \text{if $j=c$}  \\
0
      & \text{otherwise}
 \end{cases},
 \\
\Delta_3 & = &
 \begin{cases}
-h^2  \sum_{n=1}^{N} P_{cn}  P_{jn}  \theta_n
= -h^2 a_{cj}
      & \text{if $i=c$}  \\
0
      & \text{otherwise}
 \end{cases}.
\end{eqnarray*}

Thus we have
\begin{eqnarray*}
\textrm{\textbf{C}}(i,j)  =  \textrm{\textbf{C}}_1(i,j) + \textrm{\textbf{C}}_2(i,j)  = 
 \begin{cases}
2h - 2a_{cc}h^2
      & \text{if $i=j=c$}  \\
-a_{cj}h^2
      & \text{if $i=c$ and $j \neq c$} \\
-a_{ic}h^2
      & \text{if $j=c$ and $i \neq c$}
 \end{cases},
\end{eqnarray*}
and
\begin{eqnarray*}
\s\W\s^\textrm{T}  =
 \left(
  \begin{array}{ccccccc}
0 & \cdots & 0 & -a_{1c}h^2 & 0 & \cdots & 0  \\
\vdots & &\vdots & \vdots &\vdots & & \vdots \\
0 & \cdots & 0 & -a_{i-1,c}h^2 & 0 & \cdots & 0  \\
-a_{c1}h^2 & \cdots &- a_{c,i-1}h^2 & 2h-2a_{cc}h^2 & -a_{c,i+1}h^2 & \cdots & -a_{cN}h^2  \\
0 & \cdots & 0 & -a_{i+1,c}h^2 & 0 & \cdots & 0  \\
\vdots & &\vdots & \vdots& \vdots & & \vdots \\
0 & \cdots & 0 & -a_{Nc}h^2 & 0 & \cdots & 0  \\
  \end{array}
\right).
\end{eqnarray*}
As for the associated eigenvalues, we must solve the following equations
\begin{eqnarray*}
|\s\W\s^\textrm{T}  - \lambda I|  
=
 \left |
  \begin{array}{ccccccc}
-\lambda & \cdots & 0 & -a_{1c}h^2 & 0 & \cdots & 0  \\
\vdots & &\vdots & \vdots &\vdots & & \vdots \\
0 & \cdots & -\lambda & -a_{i-1,c}h^2 & 0 & \cdots & 0  \\
-a_{c1}h^2 & \cdots &- a_{c,i-1}h^2 & 2h-2a_{cc}h^2 -\lambda & -a_{c,i+1}h^2 & \cdots & -a_{cN}h^2  \\
0 & \cdots & 0 & -a_{i+1,c}h^2 & -\lambda & \cdots & 0  \\
\vdots & &\vdots & \vdots& \vdots & & \vdots \\
0 & \cdots & 0 & -a_{Nc}h^2 & 0 & \cdots & -\lambda  \\
  \end{array}
\right |
~~~~~~~~~~~~~~~~~~~~~~~~
%\nonumber \\
\end{eqnarray*}
\begin{eqnarray*}
 = 
 \left |
  \begin{array}{ccccccc}
-\lambda & \cdots & 0 & 0 & 0 & \cdots & 0  \\
\vdots & &\vdots & \vdots &\vdots & & \vdots \\
0 & \cdots & -\lambda & 0 & 0 & \cdots & 0  \\
-a_{c1}h^2 & \cdots & -a_{c,i-1}h^2 &
2h - 2a_{cc}h^2 - \lambda   + \frac{h^4}{\lambda}\sum_{i=1,i \neq c}^{N} a_{ic}a_{ci}
 &- a_{c,i+1}h^2 & \cdots & -a_{cN}h^2  \\
0 & \cdots & 0 & 0 & -\lambda & \cdots & 0  \\
\vdots & &\vdots & \vdots& \vdots & & \vdots \\
0 & \cdots & 0 & 0 & 0 & \cdots & -\lambda \\
  \end{array}
\right |
~~~~~~~~~~~~~~~~~~
%\nonumber \\
%& = &
%\left( 2h+2b_{cc}h^2 - \lambda  + \frac{h^4}{\lambda}\sum_{i=1,i \neq c}^{N} b_{ic}^2 \right) (-\lambda)^{N-1}
\nonumber \\
 = 
\left[ \lambda^2- \left( 2h-2a_{cc}h^2 \right)\lambda  - h^4\sum_{i=1,i \neq c}^{N} a_{ic}^2 \right] (-\lambda)^{N-2}
 = 
0.
~~~~~~~~~~~~~~~~~~~~~~~~~~
~~~~~~~~~~~~~~~~~~~~~~~~~~
~~~~~~~~~~~~~~~~~
\end{eqnarray*}
This yields the approximated eigenvalues as $\lambda=0$ (with multiplicity), $\lambda=h-a_{cc}h^2 $\\ $\pm \sqrt{\left( h-a_{cc}h^2 \right)^2 + h^4\sum_{i=1,i \neq c}^{N} b_{ic}^2 }$, and thus
\begin{equation*}
\lambda_{\max}=h-a_{cc}h^2 + \sqrt{h^2+2a_{cc}h^3  + h^4\sum_{i=1}^{N} a_{ic}^2 }.
\end{equation*}
Therefore, in this case, i.e., $h \rightarrow 0$, we have
\begin{equation*}
\Elb \approx \frac{1}{2\left( h-a_{cc}h^2 + \sqrt{h^2+2a_{cc}h^3  + h^4\sum_{i=1}^{N} a_{ic}^2 }\right)}.
\end{equation*}

\subsection{For large $h$}
When $h$ is large, we use the trace of $\textbf{\textrm{C}}$ to approximate its maximum eigenvalue, i.e.,
\begin{equation*}
\lambda_{\max} \approx \textrm{Tr}(\textbf{\textrm{C}}) = \sum_{i=1}^{N} \left( \textrm{\textbf{C}}_1(i,i) + \textrm{\textbf{C}}_2(i,i)  \right).
\end{equation*}
We know that
\begin{eqnarray*}
\sum_{i=1}^{N} \textrm{\textbf{C}}_1(i,i)
 & = &
\sum_{k=1}^{N}\sum_{l=1}^{N} \frac{P_{ck}P_{cl}}{\theta_k+\theta_l}\left[1- \textrm{e}^{-(\theta_k+\theta_l)h}\right]
\sum_{i=1}^{N}P_{ik}P_{il}
%\nonumber \\
 = 
\sum_{k=1}^{N} \frac{P_{ck}^2 }{2 \theta_k}\left[1- \textrm{e}^{-2 \theta_k h}\right]
%~~~~~~~~~~\left( \text{note that if $k \neq l,~~~\sum_{i=1}^{N}P_{ik}P_{il} = 0$ } \right)
%\nonumber \\
\end{eqnarray*}
\begin{eqnarray*}
& \approx &
\begin{cases}
\textrm{e}^{-2 \theta_N h}
 & \text{if $\A _1$ is Not Positive Definite (NPD)}  \\
h\sum_{k=1}^{N} P_{ck}^2 = h
& \text{if $\A _1$ is Positive Semi Definite (PSD)}   \\
\frac{1}{2}\sum_{k=1}^{N} P_{ck}^2 \theta_k ^ {-1}=  \frac{1}{2} A_1^{-1}(c,c)
& \text{otherwise, i.e., if $\A _1$ is Positve Definite (PD)}
 \end{cases},
\end{eqnarray*}
\begin{eqnarray*}
\sum_{i=1}^{N} \textrm{\textbf{C}}_2(i,i)
 & = &
\sum_{r=1}^{N} \sum_{s=1}^{N} \sum_{m=1}^{N} \sum_{n=1}^{N}
 \textrm{e}^{-( \theta_r + \theta_n ) h} P_{sr} P_{mn}
\sum_{k=1}^{N}\sum_{l=1}^{N} \frac{Q_{sk}Q_{ck}Q_{cl}Q_{ml}}{\gamma_k+\gamma_l}\left[ 1- \textrm{e}^{-(\gamma_k+\gamma_l)h} \right]
\sum_{i=1}^{N} P_{ir}P_{in}
\nonumber \\
& = &
\sum_{r=1}^{N} \sum_{s=1}^{N} \sum_{m=1}^{N}
 \textrm{e}^{-2 \theta_r  h} P_{sr} P_{mr}
\sum_{k=1}^{N}\sum_{l=1}^{N} \frac{Q_{sk}Q_{ck}Q_{cl}Q_{ml}}{\gamma_k+\gamma_l}\left[ 1- \textrm{e}^{-(\gamma_k+\gamma_l)h} \right]
\nonumber \\
& &  \left(\text{note that if $r \neq n,~~~\sum_{i=1}^{N} P_{ir}P_{in} = 0$ } \right)
\nonumber \\
& = &
\sum_{s=1}^{N} \sum_{m=1}^{N} W_2(s,m)
\textrm{e}^{-2A_1h}(s,m)
~~~~~~\left( \text{note that $\sum_{r=1}^{N} \textrm{e}^{-2 \theta_r  h} P_{sr} P_{mr} = \textrm{e}^{-2A_1h}(s,m)$ }\right )
%and $\sum_{k=1}^{N}\sum_{l=1}^{N} \frac{Q_{sk}Q_{ck}Q_{cl}Q_{ml}}{\gamma_k+\gamma_l}\left[ 1- \textrm{e}^{-(\gamma_k+\gamma_l)h} \right]= W_2(s,m) $}
\nonumber \\
& \approx &
\begin{cases}
\sum_{r=1}^{N} \sum_{s=1}^{N} \sum_{m=1}^{N}
 \textrm{e}^{-2 \theta_r  h} P_{sr} P_{mr}
& \text{if $\A _2$ is NPD or PSD} \\
\textrm{e}^{-2 \theta_N h} %(\gamma_N+\gamma_{N-1})^{-1}
& \text{if $\A _2$ is PD}
 \end{cases}.
\end{eqnarray*}

Based on the above expressions, we have
\begin{eqnarray*}
\lambda_{\max}
& \approx &
\begin{cases}
\textrm{e}^{-2 \theta_N h} + \textrm{e}^{-2 \theta_N h}  \textrm{e}^{-2\gamma_N h}
 & \text{if $\A_1$ is NPD and $\A_2$ is NPD}  \\
h + \textrm{e}^{-2\gamma_N h}
 & \text{if $\A_1$ is PSD and $\A_2$ is NPD}  \\
 \frac{1}{2} \A_1^{-1}(c,c) + \textrm{e}^{-2 \theta_N h}  \textrm{e}^{-2\gamma_N h}
 & \text{if $\A_1$ is PD and $\A_2$ is NPD}  \\
2\textrm{e}^{-2 \theta_N h}
 & \text{if $\A_1$ is NPD and $\A_2$ is PSD}  \\
h
 & \text{if $\A_1$ is PSD and $\A_2$ is PSD}  \\
\frac{1}{2} \A_1^{-1}(c,c)
 & \text{if $\A_1$ is PD and $\A_2$ is PSD}  \\
2\textrm{e}^{-2 \theta_N h}
 & \text{if $\A_1$ is NPD and $\A_2$ is PD}  \\
h+\textrm{e}^{-2 \theta_N h}
 & \text{if $\A_1$ is PSD and $\A_2$ is PD}  \\
\frac{1}{2} \A_1^{-1}(c,c) + \textrm{e}^{-2 \theta_N h}
 & \text{if $\A_1$ is PD and $\A_2$ is PD}
 \end{cases}
\end{eqnarray*}
for large $h$.

Therefore, the scaling of $\Elb$  for controlling temporal networks from $\x_0 $ to $\x_\text{f} = \textbf{\textrm{0}}$ is
\begin{eqnarray*}
\Elb
\begin{cases}
\sim h^{-1}
 ~~~ \text{small $h$}  \\
 \frac{\text{large $h$, $\A_1$ is NPD}}{\text{decreasing exponentially}}
  \begin{cases}
\sim   \textrm{e}^{2 \theta_N h}
 & \text{if $\gamma_N \geq 0$  }  \\
\sim  \textrm{e}^{2( \theta_N  + \gamma_N) h}
 & \text{if $\gamma_N < 0$ }
 \end{cases} \\
  \frac{\text{large $h$, $\A_1$ is PD}}{\text{decreasing from constant to exponentially}}
  \begin{cases}
\sim     \frac{1}{\A_1^{-1}(c,c)}
 & \text{if $\theta_N  + \gamma_N>0$}  \\
\sim  \frac{1}{2 + \A_1^{-1}(c,c)}
 & \text{if $ \theta_N  + \gamma_N  = 0$ }  \\
\sim  \textrm{e}^{2(\theta_N  + \gamma_N ) h}
& \text{if $\theta_N  + \gamma_N < 0$ }
 \end{cases} \\
   \frac{\text{large $h$, $\A_1$ is PSD}}{\text{decreasing from hyperbolically to exponentially}}
  \begin{cases}
\sim h^{-1}
& \text{if $\gamma_N \geq 0 $}  \\
\sim \textrm{e}^{2\gamma_N h}
& \text{if $\gamma_N< 0 $}
  \end{cases}
 \end{cases}.
\end{eqnarray*}

Similarly, the scaling of $\Eub$ for controlling temporal networks from $\x_0 $ to $\x_\text{f} = \textbf{\textrm{0}}$ is
\begin{eqnarray*}
\Eub
\begin{cases}
\sim h^{-N}
 ~~~ \text{small $h$}  \\
 \frac{\text{large $h$, $\A_1$ is ND}}{\text{decreasing exponentially}}
  \begin{cases}
\sim   \textrm{e}^{2 \theta_1 h}
 & \text{if $\gamma_1 \geq 0$  }  \\
\sim  \textrm{e}^{2( \theta_1  + \gamma_1) h}
 & \text{if $\gamma_1 < 0$ }
 \end{cases} \\
  \frac{\text{large $h$, $\A_1$ is NND}}{\text{decreasing from constant to exponentially}}
  \begin{cases}
\sim     C(\A_1,c)
 & \text{if $\theta_1  + \gamma_1 \geq 0$}  \\
\sim  \textrm{e}^{2(\theta_1  + \gamma_1 ) h}
& \text{if $\theta_1  + \gamma_1 < 0$ }
 \end{cases} \\
   \frac{\text{large $h$, $\A_1$ is NSD}}{\text{decreasing from hyperbolically to exponentially}}
  \begin{cases}
\sim h^{-N}
& \text{if $\gamma_1 \geq 0 $}  \\
\sim \textrm{e}^{2\gamma_1 h}
& \text{if $\gamma_1< 0 $}
  \end{cases}
 \end{cases}.
\end{eqnarray*}

\section{Control energy for $M$ snapshots and one driver node}
\label{EnergyforMnets}
When there are $M$ snapshots, we have
\begin{eqnarray*}
\s\W\s^\textrm{T}
 =
\sum_{l=1}^M \textbf{\textrm{R}}_l,
\end{eqnarray*}
where there are now $M$ terms analogous to $\textbf{\textrm{R}}_1$ and $\textbf{\textrm{R}}_2$ in the two-snapshot cases above, namely
\begin{eqnarray*}
\textbf{\textrm{R}}_M & = & \W_M
\\ \nonumber
\textbf{\textrm{R}}_l
 &=&
 \textrm{e}^{\A _{M}h_M} \cdots \textrm{e}^{\A _{l+1}h_{l+1}}
 \W_l
  \textrm{e}^{\A ^\textrm{T}_{l+1}h_{l+1}}  \cdots   \textrm{e}^{\A ^\textrm{T}_{M}h_M}
\end{eqnarray*}
for $1 \leq l \leq M-1$.

For each snapshot, we have $\A _j = \u_{j}\mathbf{\Lambda}_{j}\u_{j}^\textrm{T}$ with $ \u_{j} = \left( U^{(j)}_{r,s} \right)_{N \times N}$ and
$ \mathbf{\Lambda}_{j} = \text{diag} \left(\lambda^{(j)}_1,\lambda^{(j)}_2, \cdots, \lambda^{(j)}_N \right)$.
%$ \mathbf{\Lambda}_{j} =
%\left(
%\begin{array}{ccc}
%\lambda^{(j)}_1 & & \\
%& \ddots & \\
%& & \lambda^{(j)}_N
%  \end{array}
%\right)$.
Then we obtain
\begin{eqnarray*}
\textbf{\textrm{R}}_l
 &=&
\u_M \textrm{e}^{\Lambda_{M}h_M} \u_{M}^\textrm{T} \cdots
\u_{l+1} \textrm{e}^{\Lambda_{l+1}h_{l+1}}\u_{l+1}^\textrm{T}
 \W_l
\u_{l+1}   \textrm{e}^{\Lambda_{l+1}h_{l+1}} \u_{l+1}^\textrm{T} \cdots
\u_M \textrm{e}^{\Lambda_{M}h_M} \u_{M}^\textrm{T},
\end{eqnarray*}
with
\begin{eqnarray*}
\textbf{\textrm{R}}_l(i,j)
 &=&
\sum_{\scriptstyle i_M, O_M, \cdots, O_{l+2}, i_{l+1}, O_{l+1},
\atop
\scriptstyle O'_{l+1}, i'_{l+1}, O'_{l+2},  \cdots, O'_M,  i'_M}
U^{(M)}_{i,i_M} \textrm{e}^{\lambda^{(M)}_{i_M}h_M} U^{(M)}_{O_M,i_M} \cdots
U^{(l+1)}_{O_{l+2},i_{l+1}} \textrm{e}^{\lambda^{(l+1)}_{i_{l+1}}h_{l+1}}U^{(l+1)}_{O_{l+1},i_{l+1}}
%\\ \nonumber & &
\end{eqnarray*}
\begin{eqnarray*}
& &  \cdot
\sum_{i_l=1}^{N}\sum_{i'_l=1}^{N} \frac{U^{(l)}_{O_{l+1},i_{l}}U^{(l)}_{c,i_l}U^{(l)}_{c,i'_l}U^{(l)}_{O'_{l+1},i'_l}}{\lambda^{(l)}_{i_l}+\lambda^{(l)}_{i'_l}}\left[ \textrm{e}^{\left(\lambda^{(l)}_{i_l}+\lambda^{(l)}_{i'_l}\right)h_{l}}-1\right]
 \\ \nonumber & &
  \cdot
U^{(l+1)}_{O'_{l+1},i'_{l+1}}   \textrm{e}^{\lambda^{(l+1)}_{i'_{l+1}}h_{l+1}} U^{(l+1)}_{O'_{l+2},i'_{l+1}} \cdots
U^{(M)}_{O'_{M},i'_{M}} \textrm{e}^{\lambda^{(M)}_{i'_M}h_M} U^{(M)}_{j,i'_{M}}.
\end{eqnarray*}

For small $h_l$, we have
$$\frac{1}{\lambda^{(l)}_{i_l}+\lambda^{(l)}_{i'_l}}
\left [ \textrm{e}^{\left(\lambda^{(l)}_{i_l}+\lambda^{(l)}_{i'_l}\right)h_{l}}-1 \right] \approx h_l,$$
which leads to $\Elb \sim h^{-1}$.
% for the small duration time $h$ of each snapshot.

For large $h_l$, we have
\begin{eqnarray*}
\lambda_{\max}
 &\approx&
\sum_{l=1}^{M-1} \sum_{i=1}^{N} R_l(i,i)
+ \sum_{i=1}^{N}\sum_{i_M=1}^{N}\sum_{i'_M=1}^{N} \frac{U^{(M)}_{i,i_{M}}U^{(M)}_{c,i_M}U^{(M)}_{c,i'_M}U^{(M)}_{i,i'_M}}{\lambda^{(M)}_{i_M}+\lambda^{(M)}_{i'_M}}\left[ \textrm{e}^{\left(\lambda^{(M)}_{i_M}+\lambda^{(M)}_{i'_M}\right)h_{M}}-1\right]
\\ \nonumber &=&
\sum_{l=1}^{M-1}  \sum_{i_M=1}^N  \sum_{\scriptstyle O_M, \cdots, O_{l+2}, i_{l+1}, O_{l+1},
\atop
\scriptstyle O'_{l+1}, i'_{l+1}, O'_{l+2},  \cdots, O'_M}
 \textrm{e}^{2 \lambda^{(M)}_{i_M}h_M} U^{(M)}_{O_M,i_M} \cdots
U^{(l+1)}_{O_{l+2},i_{l+1}} \textrm{e}^{\left( \lambda^{(l+1)}_{i_{l+1}} + \lambda^{(l+1)}_{i'_{l+1}} \right) h_{l+1}}U^{(l+1)}_{O_{l+1},i_{l+1}}
\\ \nonumber & &
 \cdot
\sum_{i_l=1}^{N}\sum_{i'_l=1}^{N} \frac{U^{(l)}_{O_{l+1},i_{l}}U^{(l)}_{c,i_l}U^{(l)}_{c,i'_l}U^{(l)}_{O'_{l+1},i'_l}}{\lambda^{(l)}_{i_l}+\lambda^{(l)}_{i'_l}}\left[ \textrm{e}^{\left(\lambda^{(l)}_{i_l}+\lambda^{(l)}_{i'_l}\right)h_{l}}-1\right]
 \\ \nonumber & &
  \cdot
U^{(l+1)}_{O'_{l+1},i'_{l+1}}  U^{(l+1)}_{O'_{l+2},i'_{l+1}} \cdots
U^{(M)}_{O'_{M},i'_{M}}
+ \sum_{i_M=1}^{N}
 \frac{U^{(M)}_{c,i_{M}} U^{(M)}_{c,i_{M}} }{2 \lambda^{(M)}_{i_M}}\left[ \textrm{e}^{2 \lambda^{(M)}_{i_M}h_{M}}-1\right]
%\\ \nonumber & &
%  \\ \nonumber &=&
   \end{eqnarray*}
 \begin{eqnarray*}
& = &
 \sum_{l=1}^{M-1}
 \sum_{i_M=1}^N
 \sum_{i_l=1}^{N}
\sum_{i'_l=1}^{N}
 \sum_{\scriptstyle O_M, \cdots, O_{l+2}, i_{l+1}, O_{l+1},
\atop
\scriptstyle O'_{l+1}, i'_{l+1}, O'_{l+2},  \cdots, O'_M}
  \textrm{e}^{ 2 \lambda^{(M)}_{i_M}h_M + \sum_{s=l+1}^{M-1} \left( \lambda^{(s)}_{i_\text{s}} + \lambda^{(s)}_{i'_\text{s}} \right)h_s }
 \frac{\textrm{e}^{\left(\lambda^{(l)}_{i_l}+\lambda^{(l)}_{i'_l}\right)h_{l}}-1}{\lambda^{(l)}_{i_l}+\lambda^{(l)}_{i'_l}}U_1
 \\ \nonumber & &
+ \sum_{i_M=1}^{N}
 \frac{1 }{2 \lambda^{(M)}_{i_M}}\left[ \textrm{e}^{2 \lambda^{(M)}_{i_M}h_{M}}-1\right]U_2,
\end{eqnarray*}
where $U_1 = U^{(M)}_{O_M,i_M} \cdots U^{(l+1)}_{O_{l+2},i_{l+1}}
U^{(l+1)}_{O_{l+1},i_{l+1}}U^{(l)}_{O_{l+1},i_{l}}
U^{(l)}_{c,i_l}U^{(l)}_{c,i'_l}U^{(l)}_{O'_{l+1},i'_l}
U^{(l+1)}_{O'_{l+1},i'_{l+1}}  U^{(l+1)}_{O'_{l+2},i'_{l+1}} \cdots
U^{(M)}_{O'_{M},i'_{M}}
$ and $U_2 = U^{(M)}_{c,i_{M}} U^{(M)}_{c,i_{M}}$.

Therefore, the scaling of $\Elb$ for controlling temporal networks with $M$ snapshots from $\x_0 = \zero$ to $\x_\text{f}$ is
\begin{eqnarray*}
\Elb
\begin{cases}
\sim h^{-1}
 ~~~ \text{small $h$}  \\
 \frac{\text{large $h$, $\A_M$ is NND}}{\text{decreasing exponentially}}
  \begin{cases}
\sim   \textrm{e}^{- 2\gamma h}
 & \text{$\gamma = \max$
  $\left\{ \lambda,
  \lambda_1^{(M)} \right\}$
   }
 \end{cases} \\
 \frac{\text{large $h$, $\A_M$ is ND}}{\text{decreasing from constant to exponentially}}
  \begin{cases}
\sim     - \frac{1}{\A_M^{-1}(c,c)}
 & \text{if $\lambda < 0$}  \\
 \sim     \frac{1}{2 - \A_M^{-1}(c,c)}
 & \text{if $\lambda =0$}  \\
\sim  \textrm{e}^{- 2\lambda h}
 & \text{if $\lambda > 0$}
 \end{cases} \\
  \frac{\text{large $h$, $\A_M$ is NSD}}{\text{decreasing from hyperbolically to exponentially}}
  \begin{cases}
\sim h^{-1}
~~~ \text{if $\lambda \leq 0$ }  \\
\sim \textrm{e}^{-2 \lambda h}
 ~~~ \text{if $\lambda > 0$ }
  \end{cases}
 \end{cases},
\end{eqnarray*}
where $\lambda = \mathop {\max} \limits_{l} \left\{ \lambda_1^{(M)} + \mathop {\sum_{i=l}^{M-1}} \limits_{1 \leq l \leq M-1}  \lambda_1^{(i)} \right\}$.

And the overall scaling of $\Eub$  for controlling $M$ temporal networks from $\x_0 = \zero$ to $\x_\text{f}$ is
\begin{eqnarray*}
\Eub \sim
\begin{cases}
 h^{-N}
 ~~~ \text{small $h$}  \\
 \frac{\text{large $h$, $\A_M$ is PD}}{\text{decreasing exponentially}}
  \begin{cases}
  \textrm{e}^{- 2\gamma h}
 & \text{$\gamma = \max$
  $\left\{ \lambda,
  \lambda_N^{(M)} \right\}$
   }
 \end{cases} \\
 \frac{\text{large $h$, $\A_M$ is NPD}}{\text{decreasing from constant to exponentially}}
  \begin{cases}
    C(\A_M,c)
 & \text{if $\lambda \leq 0$}  \\
 \textrm{e}^{- 2\lambda h}
 & \text{if $\lambda > 0$}
 \end{cases} \\
  \frac{\text{large $h$, $\A_M$ is PSD}}{\text{decreasing from hyperbolically to exponentially}}
  \begin{cases}
 h^{-N}
~~~ \text{if $\lambda \leq 0$ }  \\
\textrm{e}^{-2 \lambda h}
 ~~~ \text{if $\lambda > 0$ }
  \end{cases}
 \end{cases},
\end{eqnarray*}
with $\lambda = \mathop {\max} \limits_{l} \left\{ \lambda_N^{(M)} + \mathop {\sum_{i=l}^{M-1}} \limits_{1 \leq l \leq M-1}  \lambda_N^{(i)} \right\}$.

Similarly, the scaling of $\Elb$  for controlling temporal networks from $\x_0 $ to $\x_\text{f} = \zero$ with $M$ snapshots is obtained as
\begin{eqnarray*}
\Elb \sim
\begin{cases}
 h^{-1}
 ~~~ \text{small $h$}  \\
 \frac{\text{large $h$, $\A_1$ is NPD}}{\text{decreasing exponentially}}
  \begin{cases}
   \textrm{e}^{2 \gamma h}
 & \text{$\gamma = \min$
  $\left\{ \lambda, \lambda_N^{(1)} \right\}$}
 \end{cases} \\
  \frac{\text{large $h$, $\A_1$ is PD}}{\text{decreasing from constant to exponentially}}
  \begin{cases}
     \frac{1}{\A_1^{-1}(c,c)}
 & \text{if $\lambda >0$}  \\
  \frac{1}{2 + \A_1^{-1}(c,c)}
 & \text{if $\lambda  = 0$ }  \\
  \textrm{e}^{2 \lambda h}
& \text{if $\lambda < 0$ }
 \end{cases} \\
   \frac{\text{large $h$, $\A_1$ is PSD}}{\text{decreasing from hyperbolically to exponentially}}
  \begin{cases}
 h^{-1}
& \text{if $\lambda  \geq 0 $}  \\
 \textrm{e}^{2 \lambda h}
& \text{if $\lambda  < 0 $}
  \end{cases}
 \end{cases},
\end{eqnarray*}
where $\lambda = \mathop {\min} \limits_{l} \left\{
\lambda_N^{(1)} + \mathop {\sum_{i=2}^{l}} \limits_{2 \leq l \leq M}  \lambda_N^{(i)}
\right\}$.

And the scaling of $\Eub$ for controlling temporal networks from $\x_0 $ to $\x_\text{f} = \zero$ with $M$ snapshots is
\begin{eqnarray*}
\Eub \sim
\begin{cases}
 h^{-N}
 ~~~ \text{small $h$}  \\
 \frac{\text{large $h$, $\A_1$ is ND}}{\text{decreasing exponentially}}
  \begin{cases}
   \textrm{e}^{2 \gamma h}
 & \text{$\gamma = \min$
  $\left\{ \lambda, \lambda_1^{(1)} \right\}$}
 \end{cases} \\
  \frac{\text{large $h$, $\A_1$ is NND}}{\text{decreasing from constant to exponentially}}
  \begin{cases}
     C(\A_1,c)
 & \text{if $\lambda \geq 0$}  \\
  \textrm{e}^{2 \lambda h}
& \text{if $\lambda < 0$ }
 \end{cases} \\
   \frac{\text{large $h$, $\A_1$ is NSD}}{\text{decreasing from hyperbolically to exponentially}}
  \begin{cases}
h^{-N}
& \text{if $\lambda \geq 0 $}  \\
 \textrm{e}^{2 \lambda h}
& \text{if $\lambda  < 0 $}
  \end{cases}
 \end{cases},
\end{eqnarray*}
where $\lambda = \mathop {\min} \limits_{l} \left\{
\lambda_1^{(1)} + \mathop {\sum_{i=2}^{l}} \limits_{2 \leq l \leq M}  \lambda_1^{(i)}
\right\}$.

\section{Control energy for $p$ driver nodes}
\label{si_M_p}

Here we provide the derivation of the control energy scaling for controlling temporal networks from $\x_0 = \textbf{\textrm{0}}$ to $\x_\text{f}$ with $p$ driver nodes, 
generalizing the single driver node case shown above.
Other cases with $\x_0 \neq \zero$ can be obtained based on the similar generalization from a single driver node case shown in Sec.~\ref{Energyfor2netsControl}.

Assuming that there are $p$ driver nodes with $\u(t) = (u_1(t), u_2(t), \cdots, u_p(t))^{\textrm{T}}$, and the set of nodes receiving inputs is the set $\mathcal{I} = \{ i_1, i_2, \cdots, i_p\}$.
Without loss of generality, we can relabel the network nodes so that the control inputs correspond to nodes $1,2,\cdots,p$ by letting $i_j = j$ for $j = 1, 2,\cdots, p$. 
Hence node $j$ corresponds to the input $u_j(t)$, and we have $\mathcal{I} = \{ 1, 2, \cdots, p\}$.
%(actually our methods are still available for the case where $i_j \neq j$, here we let $i_j = j$ just to let the method to be understood easier). 
%Without loss of generality, we can assume these inputs are applied to nodes $1, 2, \ldots, p$ by simply relabeling the network nodes.
Then we obtain that $B_{ii} = 1$ for $i = 1, 2,\cdots, p$, with all other entries of $\B$ equal to $0$.
We shall first consider the control energy with two snapshots, and from there generalize to an arbitrary number of snapshots.

\subsection{For two snapshots} 

In this case, the analogous terms $\textrm{\textbf{R}}_1(i,j)$ and $\textrm{\textbf{R}}_2(i,j)$  that appear in equations~(\ref{R1ij}) and~(\ref{R2ij}) of the effective gramian matrix are
%As to $\textrm{\textbf{R}}_1(i,j)$ and $\textrm{\textbf{R}}_2(i,j)$ given in equations~(\ref{R1ij}) and~(\ref{R2ij}), we have
\begin{eqnarray*}
\textrm{\textbf{R}}_1(i,j)
& = &
\sum_{r=1}^{N} \sum_{s=1}^{N} \sum_{m=1}^{N} \sum_{n=1}^{N}
Q_{ir} \textrm{e}^{( \gamma_r + \gamma_n ) h} Q_{sr} Q_{mn}  Q_{jn}
\nonumber \\ 
& &
\cdot
\sum_{k=1}^{N}\sum_{x=1}^{N}\sum_{s=1}^{p}\sum_{y=1}^{N}\sum_{l=1}^{N}
\frac{P_{sk}P_{xk}B_{xs}B_{ys}P_{yl}P_{ml}}{\theta_k+\theta_l}\left[ \textrm{e}^{(\theta_k+\theta_l)h}-1\right],
\\
\textrm{\textbf{R}}_2(i,j)
 &=&
\sum_{k=1}^{N}\sum_{x=1}^{N}\sum_{s=1}^{p}\sum_{y=1}^{N}\sum_{l=1}^{N} 
\frac{Q_{ik}Q_{xk}B_{xs}B_{ys}Q_{yl}Q_{jl}}{\gamma_k+\gamma_l}\left[ \textrm{e}^{(\gamma_k+\gamma_l)h}-1\right].
\end{eqnarray*}
%for $p$ driver nodes.
Denoting $ \B \B^{\textrm{T}}= (B'_{xy})_{N \times N}$, we have
\begin{equation*}
B'_{xy}  = 
\sum_{s=1}^{p} B_{xs}B_{ys} = 
\begin{cases}
1
      & \text{if $1 \leq x=y \leq p$}  \\
0
      & \text{otherwise}      
 \end{cases},
\end{equation*}
but by construction $ \B = \begin{pmatrix}
\textbf{\textrm{I}} \\
\textbf{\textrm{0}} 
\end{pmatrix}$ 
and therefore
$
\B \B^{\textrm{T}} = 
 \begin{pmatrix}
\textbf{\textrm{I}} \\
\textbf{\textrm{0}} 
\end{pmatrix}
  \begin{pmatrix}
\textbf{\textrm{I}}  & \textbf{\textrm{0}} 
\end{pmatrix} =
 \begin{pmatrix}
\textbf{\textrm{I}} & \textbf{\textrm{0}} \\
\textbf{\textrm{0}}  & \textbf{\textrm{0}}
\end{pmatrix}
$
where $\textbf{\textrm{I}}$ is the identity matrix with size $p$.
Hence we obtain
\begin{eqnarray*}
\textrm{\textbf{R}}_1(i,j)
& = &
\sum_{r=1}^{N} \sum_{s=1}^{N} \sum_{m=1}^{N} \sum_{n=1}^{N}
Q_{ir} \textrm{e}^{( \gamma_r + \gamma_n ) h} Q_{sr} Q_{mn}  Q_{jn}
\sum_{k=1}^{N}\sum_{c=1}^{p}\sum_{l=1}^{N}
\frac{P_{sk}P_{ck}P_{cl}P_{ml}}{\theta_k+\theta_l}\left[ \textrm{e}^{(\theta_k+\theta_l)h}-1\right],
\\
\textrm{\textbf{R}}_2(i,j)
 &=&
\sum_{k=1}^{N}\sum_{c=1}^{p}\sum_{l=1}^{N} 
\frac{Q_{ik}Q_{ck}Q_{cl}Q_{jl}}{\gamma_k+\gamma_l}\left[ \textrm{e}^{(\gamma_k+\gamma_l)h}-1\right].
\end{eqnarray*}

Then we analyze the maximum eigenvalue $\lambda_{\max}$ of $\s\W\s^\textrm{T}$ as follows.

\subsubsection{As $h \rightarrow 0$ }
In this case, we obtain
\begin{eqnarray*}
\textrm{\textbf{R}}_1(i,j)
 & \approx &
{\Omega_1} + {\Omega_2} + {\Omega_3}, ~~~~~~
\textrm{\textbf{R}}_2(i,j)
  \approx 
\begin{cases}
h
      & \text{if $i=j \in \mathcal{I}$}  \\
0
      & \text{otherwise}
 \end{cases},
\end{eqnarray*}
and 
\begin{eqnarray*}
\Omega_1  & = &
 \begin{cases}
h
      & \text{if $i=j \in \mathcal{I}$}  \\
0
      & \text{otherwise}
 \end{cases},
 \\
\Omega_2  & = &
 \begin{cases}
h^2  \sum_{r=1}^{N} Q_{ir}  Q_{jr}  \gamma_r
= h^2 b_{ij}
      & \text{if $j \in \mathcal{I}$}  \\
0
      & \text{otherwise}
 \end{cases},
 \\
\Omega_3 & = &
 \begin{cases}
h^2  \sum_{n=1}^{N} Q_{in}  Q_{jn}  \gamma_n
= h^2 b_{ij}
      & \text{if $i \in \mathcal{I}$}  \\
0
      & \text{otherwise}
 \end{cases}.
\end{eqnarray*}
%For detailed expressions, please see Eqs~(\ref{R2ijsh}) and (\ref{R1ijsh}).
Furthermore, we have
\begin{eqnarray*}
\textrm{\textbf{R}}_1(i,j) + \textrm{\textbf{R}}_2(i,j) =  
 \begin{cases}
2h+ 2b_{ii}h^2
      & \text{if $i=j \in \mathcal{I}$}  \\
b_{ij}h^2
      & \text{if $i \in \mathcal{I}$ and $j \neq i$} \\
b_{ij}h^2
      & \text{if $j \in \mathcal{I}$ and $i \neq j$}
 \end{cases},
\end{eqnarray*}
and
\begin{eqnarray*}
\s\W\s^\textrm{T} =
 \left(
  \begin{array}{cccccc}
2h + 2b_{1,1}h^2 & \cdots & b_{1,p}h^2 & b_{1,p+1}h^2 & \cdots & b_{1N}h^2  \\
\vdots & &\vdots  &\vdots & & \vdots \\
 b_{p,1}h^2  & \cdots & 2h + 2b_{p,p}h^2  & b_{p,p+1}h^2 & \cdots &  b_{p,N}h^2  \\
b_{p+1,1}h^2  & \cdots & b_{p+1,p}h^2 &  0 & \cdots & 0  \\
\vdots & &\vdots & \vdots & & \vdots \\
b_{N,1}h^2 & \cdots & b_{N,p}h^2& 0 & \cdots & 0  \\
  \end{array}
\right).
\end{eqnarray*}
As for the associated eigenvalues of $\s\W\s^\textrm{T}$, we must solve the following equations
\begin{eqnarray*}
|\s\W\s^\textrm{T} - \lambda \textrm{\textbf{I}}|  
=
 \left |
  \begin{array}{cccccc}
2h + 2b_{1,1}h^2 -\lambda & \cdots & b_{1,p}h^2 & b_{1,p+1}h^2 & \cdots & b_{1N}h^2  \\
\vdots & &\vdots  &\vdots & & \vdots \\
 b_{p,1}h^2  & \cdots & 2h + 2b_{p,p}h^2 -\lambda  & b_{p,p+1}h^2 & \cdots &  b_{p,N}h^2   \\
b_{p+1,1}h^2  & \cdots & b_{p+1,p}h^2 & -\lambda & \cdots & 0  \\
\vdots & &\vdots & \vdots & & \vdots \\
b_{N,1}h^2 & \cdots & b_{N,p}h^2& 0 & \cdots & -\lambda  \\
  \end{array} 
\right |
~~~~~~~~~~~~~~~~~~~~~~~~~~~~~~~~~~~~~~
\nonumber \\
= 
 \left |
  \begin{array}{cccccc}
2h + 2b_{1,1}h^2 -\lambda + \sum_{l = p+1}^N \frac{b_{l,1}^2h^4}{\lambda} & \cdots & b_{1,p}h^2 + \sum_{l=p+1}^{N} \frac{b_{l,p}b_{1,l}h^4}{\lambda}& 0 & \cdots & 0  \\
\vdots & &\vdots  &\vdots & & \vdots \\
 b_{p,1}h^2 + \sum_{l=p+1}^{N} \frac{b_{l,1}b_{p,l}h^4}{\lambda}  & \cdots & 2h + 2b_{p,p}h^2 -\lambda + \sum_{l = p+1}^N \frac{b_{l,p}^2h^4}{\lambda} & 0 & \cdots &  0  \\
0  & \cdots & 0 & -\lambda & \cdots & 0  \\
\vdots & &\vdots & \vdots & & \vdots \\
0 & \cdots & 0 & 0 & \cdots & -\lambda  \\
  \end{array} 
\right |
%\nonumber \\
%& = &
%\left( 2h+2b_{cc}h^2 - \lambda  + \frac{h^4}{\lambda}\sum_{i=1,i \neq c}^{N} b_{ic}^2 \right) (-\lambda)^{N-1}
~~~~~~~
\nonumber \\
= 
 \left |
  \begin{array}{cccccc}
2h + 2b_{1,1}h^2 -\lambda + \sum_{l = p+1}^N \frac{b_{l,1}^2h^4}{\lambda} & \cdots & b_{1,p}h^2 + \sum_{l=p+1}^{N} \frac{b_{l,p}b_{1,l}h^4}{\lambda} \\
\vdots & &\vdots \\
 b_{p,1}h^2 + \sum_{l=p+1}^{N} \frac{b_{l,1}b_{p,l}h^4}{\lambda}  & \cdots & 2h + 2b_{p,p}h^2 -\lambda + \sum_{l = p+1}^N \frac{b_{l,p}^2h^4}{\lambda} \\
  \end{array} 
\right |
(-\lambda)^{N-p}.
~~~~~~~~~~~~~~
\end{eqnarray*}
When $h \rightarrow 0$, the determinant of $\s\W\s^\textrm{T} - \lambda \textrm{\textbf{I}}$ can be approximated by its first-- order expression with respect to $h$, which yields $|\s\W\s^\textrm{T} - \lambda \textrm{\textbf{I}}| \approx (2h - \lambda)^p(-\lambda)^{N-p}$.
Hence we have 
$$\Elb \sim h^{-1}.$$

\subsubsection{For large $h$}
When $h$ is large, we have
\begin{eqnarray*}
\sum_{i=1}^{N} \textrm{\textbf{R}}_1(i,i)
 & \approx &
\begin{cases}
\sum_{r=1}^{N} \sum_{s=1}^{N} \sum_{m=1}^{N}
 \textrm{e}^{2 \gamma_r  h} Q_{sr} Q_{mr}
& \text{if $\A _1$ is ND or NSD} \\
\textrm{e}^{2 \gamma_1 h} \textrm{e}^{2 \theta_1h}
& \text{if $\A _1$ is NND}
 \end{cases},
 \end{eqnarray*}
\begin{eqnarray*}
\sum_{i=1}^{N} \textrm{\textbf{R}}_2(i,i)
& \approx &
\begin{cases}
- \sum_{c \in \mathcal{I}}\frac{1}{2} \A_2^{-1}(c,c)
& \text{if $\A _2$ is Negative Definite (ND)}  \\
p h
& \text{if $\A _2$ is Negative Semi Definite (NSD)}   \\
\textrm{e}^{2 \gamma_1 h}
& \text{otherwise, i.e., if $\A _2$ is Not Negative Definite (NND)}
 \end{cases}.
\end{eqnarray*}

Hence we obtain
\begin{eqnarray*}
\lambda_{\max}
& \approx &
\begin{cases}
2 \textrm{e}^{2 \gamma_1 h}
 & \text{if $\A _1$ is ND or NSD and $\A _2$ is NND}  \\
\textrm{e}^{2 \gamma_1 h} \textrm{e}^{2 \theta_1 h} + \textrm{e}^{2 \gamma_1 h}
 & \text{if $\A _1$ is NND and $\A _2$ is NND} \\
 ph
 & \text{if $\A _1$ is ND or NSD and $\A _2$ is NSD}  \\
\textrm{e}^{2 \gamma_1 h} \textrm{e}^{2 \theta_1 h} +ph
 & \text{if $\A _1$ is NND and $\A _2$ is NSD}  \\
  - \frac{1}{2} \sum_{c \in \mathcal{I}}  \A_2^{-1}(c,c)
 & \text{if $\A _1$ is ND or NSD and $\A _2$ is ND}  \\
\textrm{e}^{2 \gamma_1 h} \textrm{e}^{2 \theta_1 h}  - \frac{1}{2} \sum_{c \in \mathcal{I}} \A_2^{-1}(c,c)
 & \text{if $\A _1$ is NND and $\A _2$ is ND}
 \end{cases}
\end{eqnarray*}
for large $h$.

Therefore, the scaling of $\Elb$ for controlling temporal networks from $\x _0 = \textrm{\textbf{0}}$ to $\x_\text{f}$ with $p$ driver nodes is
\begin{eqnarray*}
\Elb \sim
\begin{cases}
 h^{-1}
 ~~~ \text{small $h$}  \\
 \frac{\text{large $h$, $\A_2$ is NND}}{\text{decreasing exponentially}}
  \begin{cases}
   \textrm{e}^{-2 \gamma_1 h}
 & \text{if $\theta_1 \leq 0$  }  \\
  \textrm{e}^{- 2(\gamma_1+\theta_1) h}
 & \text{if $\theta_1 > 0$  }
 \end{cases} \\
 \frac{\text{large $h$, $\A_2$ is ND}}{\text{decreasing from constant to exponentially}}
  \begin{cases}
    - \frac{1}{\sum_{c \in \mathcal{I}}\A_2^{-1}(c,c)}
 & \text{if $\gamma_1+\theta_1< 0$}  \\
     \frac{1}{2 - \sum_{c \in \mathcal{I}} \A_2^{-1}(c,c)}
 & \text{if $\gamma_1+\theta_1 =0$}  \\
  \textrm{e}^{- 2(\gamma_1+\theta_1) h}
 & \text{if $\gamma_1+\theta_1 > 0$}
 \end{cases} \\
  \frac{\text{large $h$, $\A_2$ is NSD}}{\text{decreasing from hyperbolically to exponentially}}
  \begin{cases}
 h^{-1}
~~~ \text{if $\theta_1 \leq 0$ }  \\
 \textrm{e}^{-2\theta_1h}
 ~~~ \text{if $\theta_1  > 0$ }
  \end{cases}
 \end{cases}.
\end{eqnarray*}

By fitting the numerical calculations, we have the scaling of $\Eub$  for controlling temporal networks from $\x _0 = \textrm{\textbf{0}}$ to $\x_\text{f}$ with $p$ driver nodes is
\begin{eqnarray*}
\Eub \sim
\begin{cases}
 h^{-N}
 ~~~ \text{small $h$}  \\
 \frac{\text{large $h$, $\A_2$ is PD}}{\text{decreasing exponentially}}
  \begin{cases}
  \textrm{e}^{-2 \gamma_N h}
 & \text{if $\theta_N \leq 0$  }  \\
 \textrm{e}^{- 2(\gamma_N+\theta_N) h}
 & \text{if $\theta_N > 0$  }
 \end{cases} \\
 \frac{\text{large $h$, $\A_2$ is NPD}}{\text{decreasing from constant to exponentially}}
  \begin{cases}
     \sum_{c \in \mathcal{I}} C(\A_2, c)
 & \text{if $\gamma_N+\theta_N \leq 0$}  \\
  \textrm{e}^{- 2(\gamma_N+\theta_N) h}
 & \text{if $\gamma_N+\theta_N > 0$}
 \end{cases} \\
  \frac{\text{large $h$, $\A_2$ is PSD}}{\text{decreasing from hyperbolically to exponentially}}
  \begin{cases}
h^{-N}
~~~ \text{if $\theta_N \leq 0$ }  \\
 \textrm{e}^{-2\theta_Nh}
 ~~~ \text{if $\theta_N  > 0$ }
  \end{cases}
 \end{cases}.
\end{eqnarray*}

\subsection{For an arbitrary number of snapshots} 

If there are $M$ snapshots, we have
\begin{eqnarray*}
R_l(i,j)
 &=&
\sum_{\scriptstyle i_M, O_M, \cdots, O_{l+2}, i_{l+1}, O_{l+1},
\atop
\scriptstyle O'_{l+1}, i'_{l+1}, O'_{l+2},  \cdots, O'_M,  i'_M}
U^{(M)}_{i,i_M} \textrm{e}^{\lambda^{(M)}_{i_M}h_M} U^{(M)}_{O_M,i_M} \cdots
U^{(l+1)}_{O_{l+2},i_{l+1}} \textrm{e}^{\lambda^{(l+1)}_{i_{l+1}}h_{l+1}}U^{(l+1)}_{O_{l+1},i_{l+1}}
\\ \nonumber & &
 \cdot
\sum_{c \in \mathcal{I}}\sum_{i_l=1}^{N}\sum_{i'_l=1}^{N} \frac{U^{(l)}_{O_{l+1},i_{l}}U^{(l)}_{c,i_l}U^{(l)}_{c,i'_l}U^{(l)}_{O'_{l+1},i'_l}}{\lambda^{(l)}_{i_l}+\lambda^{(l)}_{i'_l}}\left[ \textrm{e}^{\left(\lambda^{(l)}_{i_l}+\lambda^{(l)}_{i'_l}\right)h_{l}}-1\right]
 \\ \nonumber & &
  \cdot
U^{(l+1)}_{O'_{l+1},i'_{l+1}}   \textrm{e}^{\lambda^{(l+1)}_{i'_{l+1}}h_{l+1}} U^{(l+1)}_{O'_{l+2},i'_{l+1}} \cdots
U^{(M)}_{O'_{M},i'_{M}} \textrm{e}^{\lambda^{(M)}_{i'_M}h_M} U^{(M)}_{j,i'_{M}},
\end{eqnarray*}
for the analogous terms that contribute to the effective gramian matrix (detailed notations are given in Sec.~\ref{EnergyforMnets}).

For small $h_l$, we have
$$\frac{1}{\lambda^{(l)}_{i_l}+\lambda^{(l)}_{i'_l}}
\left [ \textrm{e}^{\left(\lambda^{(l)}_{i_l}+\lambda^{(l)}_{i'_l}\right)h_{l}}-1 \right] \approx h_l,$$
which leads to $\Elb \sim h^{-1}$ for short snapshot durations.

Conversely, for large $h_l$ we have
\begin{eqnarray*}
\lambda^{*}_{\max}
 &\approx&
 \sum_{l=1}^{M-1}
 \sum_{i_M=1}^N
 \sum_{i_l=1}^{N}
\sum_{i'_l=1}^{N}
 \sum_{\scriptstyle O_M, \cdots, O_{l+2}, i_{l+1}, O_{l+1},
\atop
\scriptstyle O'_{l+1}, i'_{l+1}, O'_{l+2},  \cdots, O'_M}
  \textrm{e}^{ 2 \lambda^{(M)}_{i_M}h_M + \sum_{s=l+1}^{M-1} \left( \lambda^{(s)}_{i_\text{s}} + \lambda^{(s)}_{i'_\text{s}} \right)h_s }
 \frac{\textrm{e}^{\left(\lambda^{(l)}_{i_l}+\lambda^{(l)}_{i'_l}\right)h_{l}}-1}{\lambda^{(l)}_{i_l}+\lambda^{(l)}_{i'_l}}U_1
% \\ \nonumber & &
\end{eqnarray*}
\begin{eqnarray*}
+ \sum_{c \in \mathcal{I}} \sum_{i_M=1}^{N}
 \frac{1 }{2 \lambda^{(M)}_{i_M}}\left[ \textrm{e}^{2 \lambda^{(M)}_{i_M}h_{M}}-1\right] U^{(M)}_{c,i_{M}} U^{(M)}_{c,i_{M}},
\end{eqnarray*}
where $U_1 = U^{(M)}_{O_M,i_M} \cdots U^{(l+1)}_{O_{l+2},i_{l+1}}
U^{(l+1)}_{O_{l+1},i_{l+1}}U^{(l)}_{O_{l+1},i_{l}}
U^{(l)}_{c,i_l}U^{(l)}_{c,i'_l}U^{(l)}_{O'_{l+1},i'_l}
U^{(l+1)}_{O'_{l+1},i'_{l+1}}  U^{(l+1)}_{O'_{l+2},i'_{l+1}} \cdots
U^{(M)}_{O'_{M},i'_{M}}
$.

Therefore, the scaling of $\Elb$ for controlling $M$ temporal networks from $\x_0 = \zero$ to $\x_\text{f}$ is
\begin{eqnarray*}
\Elb \sim
\begin{cases}
 h^{-1}
 ~~~ \text{small $h$}  \\
 \frac{\text{large $h$, $\A_M$ is NND}}{\text{decreasing exponentially}}
  \begin{cases}
   \textrm{e}^{- 2\gamma h}
 & \text{$\gamma = \max$
  $\left\{ \lambda_{\max},
  \lambda_{\max}^{(M)} \right\}$
   }
 \end{cases} \\
 \frac{\text{large $h$, $\A_M$ is ND}}{\text{decreasing from constant to exponentially}}
  \begin{cases}
    - \frac{1}{\sum_{c \in \mathcal{I}} \A_M^{-1}(c,c)}
 & \text{if $\lambda < 0$}  \\
    \frac{1}{2 - \sum_{c \in \mathcal{I}} \A_M^{-1}(c,c)}
 & \text{if $\lambda =0$}  \\
  \textrm{e}^{- 2\lambda h}
 & \text{if $\lambda > 0$}
 \end{cases} \\
  \frac{\text{large $h$, $\A_M$ is NSD}}{\text{decreasing from hyperbolically to exponentially}}
  \begin{cases}
 h^{-1}
~~~ \text{if $\lambda \leq 0$ }  \\
 \textrm{e}^{-2 \lambda h}
 ~~~ \text{if $\lambda > 0$ }
  \end{cases}
 \end{cases},
\end{eqnarray*}
and $\lambda_{\max} = \mathop {\max} \limits_{l} \left\{ \lambda_{\max}^{(M)} + \mathop {\sum_{i=l}^{M-1}} \limits_{1 \leq l \leq M-1}  \lambda_{\max}^{(i)} \right\}$.

And the scaling of $\Eub$  for controlling temporal networks with  $M$ snapshots from $\x_0 = \zero$ to $\x_\text{f}$ is
\begin{eqnarray*}
\Eub \sim 
\begin{cases}
h^{-N}
 ~~~ \text{small $h$}  \\
 \frac{\text{large $h$, $\A_M$ is PD}}{\text{decreasing exponentially}}
  \begin{cases}
   \textrm{e}^{- 2\gamma h}
 & \text{$\gamma = \max$
  $\left\{ \lambda_{\min},
  \lambda_{\min}^{(M)} \right\}$
   }
 \end{cases} \\
 \frac{\text{large $h$, $\A_M$ is NPD}}{\text{decreasing from constant to exponentially}}
  \begin{cases}
     C(A_M,c)
 & \text{if $\lambda \leq 0$}  \\
  \textrm{e}^{- 2\lambda h}
 & \text{if $\lambda > 0$}
 \end{cases} \\
  \frac{\text{large $h$, $\A_M$ is PSD}}{\text{decreasing from hyperbolically to exponentially}}
  \begin{cases}
 h^{-N}
~~~ \text{if $\lambda \leq 0$ }  \\
 \textrm{e}^{-2 \lambda h}
 ~~~ \text{if $\lambda > 0$ }
  \end{cases}
 \end{cases}
\end{eqnarray*}
with $\lambda_{\min} = \mathop {\max} \limits_{l} \left\{ \lambda_{\min}^{(M)} + \mathop {\sum_{i=l}^{M-1}} \limits_{1 \leq l \leq M-1}  \lambda_{\min}^{(i)} \right\}$.

\section{Reduction of the control energy to the static network case for identical snapshots}
\label{StatciCases}

When there is only one snapshot, say $\left(\A _\text{s},\B_\text{s}\right)$, 
%(and also can be regarded as static network), our results reduce to the case for linear time invariant systems.
the temporal network can be regarded as a static network and we show that our results reduce to the static network case.
For static networks, 
%the quadratic programming should subject to is 
the constraint that governs the minimal-energy control problem is 
$\textbf{\textrm{H}} \textbf{\textrm{c}}=\textbf{\textrm{d}}=\x_\text{f} - \text{e}^{\A _\text{s}(t_\text{f}-t_0)}\x_0$ with $\textbf{\textrm{H}}=\W_\text{s} = \int_{t_{0}}^{t_\text{f}}\text{e}^{\A _\text{s}(t_\text{f}-s)}\B_\text{s}\B_\text{s}^\textrm{T}\text{e}^{\A _\text{s}^\textrm{T}(t_\text{f}-s)}\mathrm{d}s$.
Assuming the system is controllable, $\W_\text{s}$ is nonsingular, which gives the unique solution $\textbf{\textrm{c}}=\W_\text{s}^{-1} \textbf{\textrm{d}}$.
%Hence, the optimal solution from above quadratic programming could give the optimal input as 
By plugging this into our framework above, we find that the optimal control input obeys
$\u(t)=\B^{\textrm{T}}\text{e}^{\A ^\textrm{T}(t_\text{f}-t)}\W_\text{s}^{-1}\left(\x_\text{f} - \text{e}^{\A _\text{s}(t_\text{f}-t_0)}\x_0\right)$,
and the corresponding optimal energy is
\begin{equation}
\label{restatic}
E^* (\x _0, \x_\text{f}) = \frac{1}{2} \left(\x_\text{f} - \text{e}^{\A _\text{s}(t_\text{f}-t_0)}\x_0\right)^{\textrm{T}} \W_\text{s}^{-1} \left(\x_\text{f} - \text{e}^{\A _\text{s}(t_\text{f}-t_0)}\x_0\right).
\end{equation}

%And our result for $M=1$ is same as that given in \cite{Yan2012PRL}.

Equivalently, the optimal control energy for static networks can  be obtained by considering a temporal network with $M$ \emph{identical} snapshots, \emph{i.e.}, $\A _i = \A_\text{s} $ for $i=1,2,\cdots,M$. 
In this case we have the effective gramian matrix 
\begin{eqnarray*}
%\label{temptostat}
\textbf{\textrm{SWS}}^\textrm{T}
& = &
\sum_{i=1}^{M-1}
\left(
\prod_{k=M}^{i+1}
\text{e}^{\A _{k}h_k}
\W_i
\prod_{l=i+1}^{M} \text{e}^{\A ^\textrm{T}_{l}h_l}
\right)
+\W_M
\nonumber \\
&= &
\sum_{i=1}^{M-1}
\left(
\text{e}^{\A_\text{s}    \sum_{k=i+1}^M h_k}
\int_{0}^{h_i}\text{e}^{\A_\text{s}  \tau }\B\B^\textrm{T}
\text{e}^{\A_\text{s} ^\textrm{T} \tau }\mathrm{d} \tau
\text{e}^{\A_\text{s} ^\textrm{T} \sum_{l=i+1}^M   h_l}
\right)
+\W_M
\nonumber \\
&= &
\sum_{i=1}^{M-1}
\int_{0}^{h_i}\text{e}^{\A_\text{s}  (\tau+\sum_{k=i+1}^M h_k) }\B\B^\textrm{T}
\text{e}^{\A_\text{s} ^\textrm{T} (\tau+\sum_{k=i+1}^M h_k) }\mathrm{d} \tau
+\W_M
\nonumber \\
&= &
\sum_{i=1}^{M-1}
\int_{\sum_{k=i+1}^M h_k}^{\sum_{k=i}^M h_k}
\text{e}^{\A_\text{s}  s }\B\B^\textrm{T}
\text{e}^{\A_\text{s} ^\textrm{T} s }\mathrm{d} s
+\int_{0}^{h_M}\text{e}^{\A_\text{s}  \tau }\B\B^\textrm{T}\text{e}^{\A_\text{s} ^\textrm{T} \tau }\mathrm{d} \tau
\nonumber \\
&= &
\int_{0}^{\sum_{k=1}^M h_k}\text{e}^{\A_\text{s}  \tau }\B\B^\textrm{T}\text{e}^{\A_\text{s} ^\textrm{T} \tau }\mathrm{d} \tau
\nonumber \\
&= &
\W_\text{s}.
\end{eqnarray*}
Thus in either view, the energy for controlling temporal networks recapitulates the known result for static networks \cite{Yan2012PRL}.
Indeed, for controlling a static network from $\x _0 = \zero$ to $\x_\text{f}$ 
our results indicate that the energy is bounded below by
%with $M$ snapshots, we can get the form of the scaling behavior of the energy from the stimulus for $M=1$ in the case of temporal networks (see Secs.~\ref{Energyfor2netsReach} and \ref{Energyfor2netsControl}) as
\begin{eqnarray*}
\Elb
& \sim &
\begin{cases}
h^{-1}
& \text{small $h$}  \\
- \frac{1}{\A _\text{s}^{-1}(c,c)}
& \text{large $h$, $\A _\text{s}$ is ND}  \\
h^{-1}
& \text{large $h$, $\A _\text{s}$ is NSD}   \\
\textrm{e}^{-2 \delta_1 Mh}
& \text{large $h$, $\A _\text{s}$ is NND}
 \end{cases},
\end{eqnarray*}
and bounded above by 
\begin{eqnarray*}
\Eub
& \sim &
\begin{cases}
h^{-2N}
& \text{small $h$}  \\
C(\A _\text{s}, c)
& \text{large $h$, $\A _\text{s}$ is NPD}  \\
h^{-2N}
& \text{large $h$, $\A _\text{s}$ is PSD}   \\
\textrm{e}^{-2 \delta_N M h}
& \text{large $h$, $\A _\text{s}$ is PD}
 \end{cases}
\end{eqnarray*}
where $\delta_1$ and $\delta_N$ are the maximum and minimum eigenvalues of $\A _\text{s}$, respectively.
Similarly, for controlling a static network from $\x _0 $ to $\x_\text{f} = \textrm{\textbf{0}}$, the optimal energy bounds obey 
\begin{eqnarray*}
\Elb
& \sim &
\begin{cases}
h^{-1}
& \text{small $h$}  \\
\frac{1}{\A _\text{s}^{-1}(c,c)}
& \text{large $h$, $\A _\text{s}$ is PD}  \\
h^{-1}
& \text{large $h$, $\A _\text{s}$ is PSD}   \\
\textrm{e}^{2 \delta_N Mh}
& \text{large $h$, $\A _\text{s}$ is NPD}
 \end{cases},
\end{eqnarray*}
and
\begin{eqnarray*}
\Eub
& \sim &
\begin{cases}
h^{-2N}
& \text{small $h$}  \\
C(\A _\text{s}, c)
& \text{large $h$, $\A _\text{s}$ is NND}  \\
h^{-2N}
& \text{large $h$, $\A _\text{s}$ is NSD}   \\
\textrm{e}^{2 \delta_1 Mh}
& \text{large $h$, $\A _\text{s}$ is ND}
 \end{cases}.
\end{eqnarray*}

Taken together, we know that scaling behavior of the control energy for temporal networks reduces to the static network case as shown in \cite{Yan2012PRL} when all snapshots are identical.

\section{Relation between the energy needed to control static networks and the Laplacian matrix}

Because we employ the Laplacian matrix with entries $a_{ij} > 0$ for tuning $\lambda_1^{(m)}$ and $a_{ij} < 0$ for tuning $\lambda_N^{(m)}$, we get that the energy scales as
\begin{eqnarray*}
\Elb
& \sim &
\begin{cases}
h^{-1}
& \text{small $h$}  \\
- \frac{1}{\A_\text{s}^{-1}(c,c)}
& \text{large $h$, $ \lambda_1 < 0$}  \\
h^{-1}
& \text{large $h$, $\lambda_1= 0$}   \\
\textrm{e}^{-2 \lambda_1 h}
& \text{large $h$, $\lambda_1> 0$}
 \end{cases},
\end{eqnarray*}
and
\begin{eqnarray*}
\Eub
& \sim &
\begin{cases}
h^{-2N}
& \text{small $h$}  \\
C(\A_\text{s}, c)
& \text{large $h$, $\lambda_N < 0$ }  \\
h^{-2N}
& \text{large $h$, $\lambda_N = 0$}   \\
\textrm{e}^{-2 \lambda_N h}
& \text{large $h$, $\lambda_N > 0$}
 \end{cases},
\end{eqnarray*}
as controlling the static network from $\x _0 = \zero$ to $\x_\text{f}$, where $\lambda_1 = \sum_{m=1}^M \lambda_1^{(m)}$, $\lambda_N = \sum_{m=1}^M \lambda_N^{(m)}$, and the node $c$ receives the control input directly.
Similarly, controlling from $\x _0 $ to $\x_\text{f} = \zero$, we have
\begin{eqnarray*}
\Elb
& \sim &
\begin{cases}
h^{-1}
& \text{small $h$}  \\
\frac{1}{\A_\text{s}^{-1}(c,c)}
& \text{large $h$, $\lambda_N> 0$}  \\
h^{-1}
& \text{large $h$, $\lambda_N= 0$}   \\
\textrm{e}^{2 \lambda_N h}
& \text{large $h$, $\lambda_N< 0$}
 \end{cases},
\end{eqnarray*}
and
\begin{eqnarray*}
\Eub
& \sim &
\begin{cases}
h^{-2N}
& \text{small $h$}  \\
C(\A_\text{s}, c)
& \text{large $h$, $\lambda_1 > 0$}  \\
h^{-2N}
& \text{large $h$, $\lambda_1 = 0$}   \\
\textrm{e}^{2 \lambda_1 h}
& \text{large $h$, $\lambda_1 < 0$}
 \end{cases}.
\end{eqnarray*}

\section{The first and last snapshots determine the scaling behavior of the control energy}
\label{si_firstandlastSub}
The total energy for controlling temporal network from $\x _0$ to $\x_\text{f}$ can be viewed as the summation of the energy over each snapshot (given in Eq.~(\ref{restatic})), \emph{i.e.}
\begin{eqnarray}
E(\x_0,\x_\text{f}) &=& \frac{1}{2}
\sum_{i=1}^{i = M} \left(\x _{i}- \textrm{e}^{\A_{i}h}\x _{i-1} \right)^{\textrm{T}} \W_{i}^{-1} \left(\x _{i}- \textrm{e}^{\A_{i}h}\x _{i-1} \right)
\nonumber \\ &=&
\frac{1}{2} \sum_{i=1}^{i = M} \left( 
\x_i^{\textrm{T}}\W_{i}^{-1}\x_i - 2 \x_i^{\textrm{T}}\W_{i}^{-1} \textrm{e}^{\A_{i}h}\x _{i-1}
+ \x _{i-1}^{\textrm{T}}\textrm{e}^{\A_{i}^{\textrm{T}}h}\W_{i}^{-1} \textrm{e}^{\A_{i}h}\x _{i-1}
\right),
\label{totalenergy}
 \end{eqnarray}
where $\x_{i-1}$ and $\x_{i}$ are the initial and final states over the snapshot $\A_i$, and $h$ is the duration time for each snapshot.

When $E(\x_0,\x_\text{f})$ reaches its minimum value $E^*(\x_0,\x_\text{f})$, the series of intermediate states---$\x(t)$ at times $t=mh$ when the network structure changes ($m=1,2,\cdots,m-1$)---should satisfy
\begin{eqnarray*}
\frac{\partial E(\x_0,\x_\text{f})}{\partial \x_i} &=&
\frac{\partial  \left(
\x_i^{\textrm{T}}\W_{i}^{-1}\x_i - 2 \x_i^{\textrm{T}}\W_{i}^{-1} \textrm{e}^{\A_{i}h}\x _{i-1}
+ \x _{i-1}^{\textrm{T}}\textrm{e}^{\A_{i}^{\textrm{T}}h}\W_{i}^{-1} \textrm{e}^{\A_{i}h}\x _{i-1}
\right)
}{2 \partial \x_i}
\nonumber \\ & &
+ 
\frac{\partial  \left(
\x_{i+1}^{\textrm{T}}\W_{i+1}^{-1}\x_{i+1} - 2 \x_{i+1}^{\textrm{T}}\W_{i+1}^{-1} \textrm{e}^{\A_{i+1}h}\x _{i}
+ \x _{i}^{\textrm{T}}\textrm{e}^{\A_{i+1}^{\textrm{T}}h}\W_{i+1}^{-1} \textrm{e}^{\A_{i+1}h}\x _{i}
\right)
}{2 \partial \x_i}
\nonumber \\ &=&
\W_{i}^{-1}\x_i
- \W_{i}^{-1} \textrm{e}^{\A_{i}h}\x_{i-1}
-  \textrm{e}^{\A_{i+1}^{\textrm{T}}h} \W_{i+1}^{-1} \x_{i+1}
+  \textrm{e}^{\A_{i+1}^{\textrm{T}}h} \W_{i+1}^{-1} \textrm{e}^{\A_{i+1}h}\x_{i}
\nonumber \\ &=&
\zero
 \end{eqnarray*}
 for $i=1,2,\cdots,M-1$.
From the above equation, we know $E(\x_0,\x_\text{f})=E^*(\x_0,\x_\text{f})$ when the following relation holds
 \begin{eqnarray*}
\p
 \left(
  \begin{array}{c}
\x_1\\
\vdots\\
\x_{i-1} \\
\x_i\\
\x_{i+1} \\
\vdots \\
\x_{M-1}\\
  \end{array}
\right)
& = &
\left(
  \begin{array}{c}
\W_{1}^{-1} \textrm{e}^{\A_{1}h}\x_{0}  \\
\vdots \\
0 \\
0 \\
0 \\
\vdots \\
\textrm{e}^{\A_{M}^\textrm{T}h}\W_M^{-1} \x_\text{f} \\
  \end{array}
\right),
\end{eqnarray*}
where
\footnotesize
 \begin{eqnarray*}
\p =
 \left(
  \begin{array}{cccccc}
\W_{1}^{-1}  +  \textrm{e}^{\A_{2}^{\textrm{T}}h} \W_{2}^{-1} \textrm{e}^{\A_{2}h}   &
-  \textrm{e}^{\A_{2}^{\textrm{T}}h} \W_{2}^{-1}  ~~~~~~~\cdots & & \\
      &\vdots& \vdots &    \\
 \cdots ~~~  - \W_{i}^{-1} \textrm{e}^{\A_{i}h}  &
  \W_{i}^{-1}  +  \textrm{e}^{\A_{i+1}^{\textrm{T}}h} \W_{i+1}^{-1} \textrm{e}^{\A_{i+1}h} &
   -  \textrm{e}^{\A_{i+1}^{\textrm{T}}h} \W_{i+1}^{-1} ~~~~~~~~~~ \cdots & \\
   & \vdots & \vdots &     \\
    &  &  \cdots ~~~ - \W_{M-1}^{-1} \textrm{e}^{\A_{M-1}h}
    &\W_{M-1}^{-1} + \textrm{e}^{\A_{M}^\textrm{T}h} \W_{M}^{-1} \textrm{e}^{\A_{M}h} \\
  \end{array}
\right).
%\left(
%  \begin{array}{c}
%x_1\\
%\vdots\\
%x_{i-1} \\
%x_i\\
%x_{i+1} \\
%\vdots \\
%x_{M-1}\\
%  \end{array}
%\right)
% \\ \nonumber
% & = &
% P
% \left(
%  \begin{array}{c}
%x_1\\
%\vdots\\
%x_{i-1} \\
%x_i\\
%x_{i+1} \\
%\vdots \\
%x_{M-1}\\
%  \end{array}
%\right)
% \\ \nonumber
%& = &
%\left(
%  \begin{array}{c}
% \left( W_{1}^{-1} \textrm{e}^{A_{1}h}x_{0} \right) ^ \textrm{T},
%\cdots,
%0,
%0,
%0,
%\cdots,
% \left(    \textrm{e}^{A_{M}^\textrm{T}h}W_M^{-1} x_\text{f}  \right) ^ \textrm{T}
%  \end{array}
%\right) ^{\textrm{T}}
\end{eqnarray*}

\normalsize
\begin{eqnarray*}
\text{Equation~(\ref{totalenergy}) can also be written as~}
E^*(\x_0,\x_\text{f}) = \frac{1}{2} \left(\x_0^\textrm{T}, \cdots, \x_\text{f}^\textrm{T} \right) \f
\left(
\begin{array}{c}
\x_0\\
\vdots \\
 \x_\text{f}\\
  \end{array}
\right)
\text{with the following expression}
\end{eqnarray*}
\footnotesize
\begin{eqnarray*}
%\hspace{-3in} 
\f = ~~~~~~~~~~~~~~~~~~~~~~~~~~~~~~~~~~~~~~~~~~~~~~~~~~~~~~~~~~~~~~~~~~~~~~~~~~~~~~~~~~~~~~~~~~~
~~~~~~~~~~~~~~~~~~~~~~~~~~~~~~~~~~~~~~~~~~~~~~~~~~~~~~~~~~
\nonumber \\
 \left(
  \begin{array}{ccccc}
\textrm{e}^{\A_{1}^{\textrm{T}}h} \W_{1}^{-1} \textrm{e}^{\A_{1}h}    & -\textrm{e}^{\A_{1}^{\textrm{T}}h} \W_{1}^{-1}  & 0& \cdots &  \\
-\W_{1}^{-1} \textrm{e}^{\A_{1}h}
&\W_{1}^{-1} + \textrm{e}^{\A_{2}^{\textrm{T}}h} \W_{2}^{-1} \textrm{e}^{\A_{2}h}
& -\textrm{e}^{\A_{2}^{\textrm{T}}h} \W_{2}^{-1}  & \cdots& \\
&  & \vdots & &  \\
\cdots & -\W_{i}^{-1} \textrm{e}^{\A_{i}h} &\W_{i}^{-1} + \textrm{e}^{\A_{i+1}^{\textrm{T}}h} \W_{i+1}^{-1} \textrm{e}^{\A_{i+1}h}
& -\textrm{e}^{\A_{i+1}^{\textrm{T}}h} \W_{i+1}^{-1}  &  \cdots \\
&  & \vdots & &  \\
&  \cdots &   -\W_{M-1}^{-1} \textrm{e}^{\A_{M-1}h} &
\W_{M-1}^{-1} + \textrm{e}^{\A_{M}^{\textrm{T}}h} \W_{M}^{-1} \textrm{e}^{\A_{M}h} & -\textrm{e}^{\A_{M}^{\textrm{T}}h} \W_{M}^{-1}  \\
&  & \cdots &   -\W_{M}^{-1} \textrm{e}^{\A_{M}h} & \W_{M}^{-1}  \\
  \end{array}
\right)
%\nonumber \\ 
\end{eqnarray*}
\begin{eqnarray*}
 = 
 \left(
  \begin{array}{ccccc}
\textrm{e}^{\A_{1}^{\textrm{T}}h} \W_{1}^{-1} \textrm{e}^{\A_{1}h}    & -\textrm{e}^{\A_{1}^{\textrm{T}}h} \W_{1}^{-1}  & \cdots & &  \\
-\W_{1}^{-1} \textrm{e}^{\A_{1}h} & \cdots & &   & \\
\vdots &  & \p & & \vdots \\
&  & &\cdots & -\textrm{e}^{\A_{M}^{\textrm{T}}h} \W_{M}^{-1}  \\
&  & \cdots  &   -\W_{M}^{-1} \textrm{e}^{\A_{M}h} & \W_{M}^{-1}  \\
  \end{array}
\right).
~~~~~~~~~~~~~~~~~~~~~~~~~~~~~~~~~~~~~~~~~~~~~~~~~~~~~~
%\left(
%  \begin{array}{c}
%x_1\\
%\vdots\\
%x_{i-1} \\
%x_i\\
%x_{i+1} \\
%\vdots \\
%x_{M-1}\\
%  \end{array}
%\right)
% \\ \nonumber
%=
%\left(
%  \begin{array}{c}
% W_{1}^{-1} \textrm{e}^{A_{1}h}x_{0},
%\cdots,
%0,
%0,
%0,
%\cdots,
%\textrm{e}^{A_{M}^\textrm{T}h}W_M^{-1} x_\text{f}
%  \end{array}
%\right) ^{\textrm{T}}
\end{eqnarray*}
\normalsize
Considering that
$$\left(
  \begin{array}{ccc}
\x_1^\textrm{T},
\cdots,
\x_{M-1}^\textrm{T}
  \end{array}
\right)
\p
 \left(
  \begin{array}{c}
\x_1 \\
\vdots\\
\x_{M-1}\\
  \end{array}
\right)
=
\x_1^\textrm{T}\W_{1}^{-1} \textrm{e}^{\A_{1}h}\x_{0}  +  \x_{M-1}^\textrm{T}\textrm{e}^{\A_{M}^\textrm{T}h}\W_M^{-1} \x_\text{f},$$ we further have
\begin{eqnarray*}
2E &=&
\x_0^{\textrm{T}} \textrm{e}^{\A_{1}^{\textrm{T}}h} \W_{1}^{-1} \textrm{e}^{\A_{1}h} \x_0 -
\x_0^{\textrm{T}} \textrm{e}^{\A_{1}^{\textrm{T}}h} \W_{1}^{-1} x_1 -
\x_1 ^{\textrm{T}} \W_{1}^{-1} \textrm{e}^{\A_{1}h} \x_0
\\ \nonumber
&&- \x_{M-1} ^{\textrm{T}} \textrm{e}^{\A_{M}^{\textrm{T}}h} \W_{M}^{-1} \x_M -
\x_{M} ^{\textrm{T}} \W_{M}^{-1} \textrm{e}^{\A_{M}h} \x_{M-1} +
\x_{M} ^{\textrm{T}}  \W_{M}^{-1} \x_M
\\ \nonumber
&& +
\left(
  \begin{array}{ccc}
\x_1^\textrm{T},
\cdots,
\x_{M-1}^\textrm{T}
  \end{array}
\right)
\p
 \left(
  \begin{array}{c}
\x_1 \\
\vdots\\
\x_{M-1}\\
  \end{array}
\right)
\\ \nonumber
&=&
\x_0^{\textrm{T}} \textrm{e}^{\A_{1}^{\textrm{T}}h} \W_{1}^{-1} \textrm{e}^{\A_{1}h} \x_0 -
\x_0^{\textrm{T}} \textrm{e}^{\A_{1}^{\textrm{T}}h} \W_{1}^{-1} \x_1
- \x_{M-1} ^{\textrm{T}} \textrm{e}^{\A_{M}^{\textrm{T}}h} \W_{M}^{-1} \x_M +
\x_{M} ^{\textrm{T}}  \W_{M}^{-1} \x_M.
\end{eqnarray*}
Hence we obtain
\begin{eqnarray*}
E^*(\x_0,\x_\text{f}) =
\frac{1}{2}
\left( \x_0^{\textrm{T}} \textrm{e}^{\A_{1}^{\textrm{T}}h} \W_{1}^{-1} \textrm{e}^{\A_{1}h} \x_0 -
\x_0^{\textrm{T}} \textrm{e}^{\A_{1}^{\textrm{T}}h} \W_{1}^{-1} \x_1
- \x_{M-1} ^{\textrm{T}} \textrm{e}^{\A_{M}^{\textrm{T}}h} \W_{M}^{-1} \x_M +
\x_{M} ^{\textrm{T}}  \W_{M}^{-1} \x_M
\right).
\end{eqnarray*}
From the last expression, we find that the optimal energy has no direct dependence on the intermediate states except $\x_1$ and $\x_{M-1}$, 
suggesting that it is the properties of $\A_1$ and $\A_M$ have the most influence on $E^*(\x_0,\x_\text{f})$.
%meaning that as we analyze the properties of $E^*(\x_0,\x_\text{f})$ more attention should be put on $\A_1$ and $\A_M$.
%But, at the same time, one should note that the optimal values of $\x_1$ and $\x_{M-1}$ under the minimum control energy are got from all $\A_i$.

\section{Supplementary Tables and Figures}

%\newpage
%\begin{landscape}
\begin{table} %\footnotesize
\caption{
Scaling behavior of $\Elb$ for a temporal network and its static counterpart for $\x_0 = \zero$ and $\x_\text{f}$ with $5$ snapshots.
We summarize the theoretical scaling behavior from results for large $h$ as $M = 5$ in this table.
$\lambda_1^{(i)}$ is the maximum eigenvalue of the snapshot $A_i$ with $i = 1, 2 ,3, 4, 5$.
$\gamma, \delta$, and $\lambda$
% (the expression can be found in equation (?)) 
determine the energy scaling behavior for controlling temporal network $\Elb_\text{t}$ and static network $\Elb_\text{s}$.
We employ $\lambda_1^{(5)} = 2, -2, \textrm{and} ~0$ to represent the Not Negative, Negative, and Negative Semi definite of the last snapshot matrix $\A _5$, respectively.
%$\Elb_\textrm{t}$ ($\Elb_\textrm{s}$) corresponds to temporal (static) networks.
For the constant numbers, $C_1 = -1/\A_\text{s} ^{-1}(c,c)$, $C_2 = 1/(2-\A _5^{-1}(c,c))$, and $C_3 = -1/\A _5^{-1}(c,c)$, where $\A_\text{s}$ and $\A_5$ correspond to the network matrix in each case.
Here $N = 8$, $k = 4$, and the node receiving the input directly is node $3$.
For each setting of the $\lambda_1^{(i)}$, the numerical validation of our theoretical result is presented in Fig.~\ref{figsi_min2}, and here we indicate the corresponding panel.
}
\center
\begin{tabular}{>{\small}c>{\small}c>{\small}c>{\small}c|>{\small}c>{\small}c|>{\small}c>{\small}c>{\small}c|>{\small}c>{\small}c|>{\small}c>{\small}c>{\small}c|>{\small}c>{\small}c|>{\small}c>{\small}c>{\small}c}
\hline \hline
 \multicolumn{4}{c|}{Maximum eigenvalues} &  \multicolumn{2}{c}{$\lambda_1^{(5)} = 2$}  &  \multicolumn{3}{c|}{Scaling}  & \multicolumn{2}{c}{$\lambda_1^{(5)} = -2$} &  \multicolumn{3}{c|}{Scaling}     &  \multicolumn{2}{c}{$\lambda_1^{(5)} = 0$} &  \multicolumn{3}{c}{Scaling} \\
$\lambda_1^{(1)}$  & $\lambda_1^{(2)}$ & $\lambda_1^{(3)}$ & $\lambda_1^{(4)}$ &
$\gamma$  &   \multicolumn{1}{c}{$\delta$}  &    $\Elb_\textrm{t}$  &  $\Elb_\textrm{s}$  &  Panel  &
$\lambda$  &   \multicolumn{1}{c}{$\delta$}   &  $\Elb_\textrm{t}$  &  $\Elb_\textrm{s}$  &  Panel  &
$\lambda$  &    \multicolumn{1}{c}{$\delta$}   & $\Elb_\textrm{t}$  &  $\Elb_\textrm{s}$  & Panel    \\  \hline
2& 2& 2& 2&      10& 10&   $\textrm{e}^{-20}$   & $\textrm{e}^{-20}$ & (a) & 6& 6&     $\textrm{e}^{-12}$& $\textrm{e}^{-12}$& (e)&    8& 8&  $\textrm{e}^{-16}$& $\textrm{e}^{-16}$& (i) \\
-2& 2& 2& 2&            8& 6&   $\textrm{e}^{-16}$   & $\textrm{e}^{-12}$ & (a)&  6& 2  & $\textrm{e}^{-12}$& $\textrm{e}^{-4}$& (e)& 6& 4&  $\textrm{e}^{-12}$   & $\textrm{e}^{-8}$& (i)  \\
2& -2& 2& 2&               6& 6&   $\textrm{e}^{-12}$   & $\textrm{e}^{-12}$& (a) & 2& 2&   $\textrm{e}^{-4}$& $\textrm{e}^{-4}$& (e)& 4& 4&  $\textrm{e}^{-8}$& $\textrm{e}^{-8}$ & (i)\\
2& 2& -2& 2&            6& 6&      $\textrm{e}^{-12}$   & $\textrm{e}^{-12}$ & (a)&  2& 2&   $\textrm{e}^{-4}$& $\textrm{e}^{-4}$& (e)& 4& 4 & $\textrm{e}^{-8}$& $\textrm{e}^{-8}$& (i) \\
2& 2& 2& -2&         6& 6&    $\textrm{e}^{-12}$   & $\textrm{e}^{-12}$& (b) &  2& 2&    $\textrm{e}^{-4}$& $\textrm{e}^{-4}$& (f)& 4& 4 & $\textrm{e}^{-8}$& $\textrm{e}^{-8}$& (j)\\
-2& -2& 2& 2&           6& 2&   $\textrm{e}^{-12}$   & $\textrm{e}^{-4}$ & (b)&  2& -2&    $\textrm{e}^{-4}$& $C_1$& (f)&  4& 0 & $\textrm{e}^{-8}$& $h^{-1}$& (j)\\
-2& 2& -2& 2&           4& 2&  $\textrm{e}^{-8}$   & $\textrm{e}^{-4}$& (b) & 0& -2&    $ C_2$& $C_1$& (f)&  2& 0& $\textrm{e}^{-4}$& $h^{-1}$& (j)\\
-2& 2& 2& -2&           4& 2&   $\textrm{e}^{-8}$   & $\textrm{e}^{-4}$& (b) & 0& -2&   $ C_2$& $C_1$& (f)& 2& 0& $\textrm{e}^{-4}$& $h^{-1}$& (j) \\
2& -2& -2& 2&         4& 2& $\textrm{e}^{-8}$   & $\textrm{e}^{-4}$ & (c) &  0 & -2&   $ C_2 $& $C_1$& (g)&   2& 0& $\textrm{e}^{-4}$& $h^{-1}$& (k) \\
2& -2& 2& -2&      2& 2&   $\textrm{e}^{-4}$   & $\textrm{e}^{-4}$& (c)& -2& -2&   $C_3$& $C_1$& (g)& 0& 0& $h^{-1}$& $h^{-1}$& (k) \\
2& 2& -2& -2&           2& 2&    $\textrm{e}^{-4}$   & $\textrm{e}^{-4}$& (c) &  -2& -2&     $C_3$& $C_1$& (g)& 0& 0 & $h^{-1}$& $h^{-1}$& (k) \\
-2& -2& -2& 2&           4& -2&     $\textrm{e}^{-8}$   & $C_1$ & (c) & 0& -6&    $C_2$& $C_1$& (g)&  2& -4& $\textrm{e}^{-4}$& $C_1$& (k)  \\
-2& -2& 2& -2&               2& -2&        $\textrm{e}^{-4}$   & $C_1$& (d)& -2& -6&  $C_3$& $C_1$& (h)&  0& -4 & $h^{-1}$& $C_1$ & (l)\\
-2& 2& -2& -2&           2& -2&        $\textrm{e}^{-4}$   & $C_1$& (d)& -4& -6&     $C_3$& $C_1$& (h)& -2& -4& $h^{-1}$& $C_1$& (l) \\
2& -2& -2& -2&           2& -2&      $\textrm{e}^{-4}$   & $C_1$& (d)& -6& -6&    $C_3$& $C_1$& (h)&  -4& -4 & $h^{-1}$& $C_1$& (l) \\
-2& -2& -2& -2&          2 & -6 &      $\textrm{e}^{-4}$   & $C_1$& (d) &  -4 & -10 &   $C_3$ & $C_1$& (h) & -2 & -8 & $h^{-1}$ & $C_1$ & (l) \\
\hline \hline
\end{tabular}
\label{minimumenergyforM}
\end{table}

\begin{table} % \footnotesize
\caption{
Scaling behavior of $\Eub$ for a temporal network and its static counterpart for $\x_0 = \zero$ and $\x_\text{f}$ with $5$ snapshots.
We summarize the theoretical scaling behavior from results for large $h$ as $M = 5$ in this table.
$\lambda_N^{(i)}$ is the minimum eigenvalue of snapshot $A_i$ with $i = 1, 2 ,3, 4, 5$.
$\gamma,~\delta$, and $\lambda$ determine the scaling behavior for controlling temporal network $\Eub_\textrm{t}$ and static network $\Eub_\textrm{s}$.
We employ $\lambda_1^{(5)} = 2, -2, \textrm{and} ~0$ to represent the Positive, Not Positive, and Positive Semi definite of the last snapshot matrix $A_5$, respectively.
%$\Eub_\textrm{t}$ ($\Eub_\textrm{s}$) corresponds to temporal (static) networks.
For the constant numbers, $C_1 = -1/\A_\text{s} ^{-1}(c,c)$, $C_2 = 1/(2-\A _5^{-1}(c,c))$, and $C_3 = -1/\A _5^{-1}(c,c)$, where $\A_\text{s}$ and $\A_5$ correspond to the network matrix in each case.
Here $N = 8$, $k = 4$, and the node receiving the input directly is node $5$.
For each setting of the $\lambda_N^{(i)}$, the numerical validation of our theoretical result is presented in Fig.~\ref{figsi_max2}, and here we indicate the corresponding panel.
}
\center
\begin{tabular}{>{\small}c>{\small}c>{\small}c>{\small}c|>{\small}c>{\small}c|>{\small}c>{\small}c>{\small}c|>{\small}c>{\small}c|>{\small}c>{\small}c>{\small}c|>{\small}c>{\small}c|>{\small}c>{\small}c>{\small}c}
\hline \hline
 \multicolumn{4}{c|}{Minimum eigenvalues} &  \multicolumn{2}{c}{$\lambda_N^{(5)} = 2$}  &  \multicolumn{3}{c|}{Scaling}  & \multicolumn{2}{c}{$\lambda_N^{(5)} = -2$} &  \multicolumn{3}{c|}{Scaling}     &  \multicolumn{2}{c}{$\lambda_N^{(5)} = 0$} &  \multicolumn{3}{c}{Scaling} \\
$\lambda_N^{(1)}$  & $\lambda_N^{(2)}$ & $\lambda_N^{(3)}$ & $\lambda_N^{(4)}$ &
$\gamma$  &   \multicolumn{1}{c}{$\delta$}  &    $\Eub_\textrm{t}$  &  $\Eub_\textrm{s}$  &  Panel  &
$\lambda$  &   \multicolumn{1}{c}{$\delta$}   &  $\Eub_\textrm{t}$  &  $\Eub_\textrm{s}$  &  Panel  &
$\lambda$  &    \multicolumn{1}{c}{$\delta$}   & $\Eub_\textrm{t}$  &  $\Eub_\textrm{s}$  & Panel    \\  \hline
2& 2& 2& 2&      10& 10&   $\textrm{e}^{-20}$   & $\textrm{e}^{-20}$ & (a) & 6& 6&     $\textrm{e}^{-12}$& $\textrm{e}^{-12}$& (e)&    8& 8&  $\textrm{e}^{-16}$& $\textrm{e}^{-16}$& (i) \\
-2& 2& 2& 2&            8& 6&   $\textrm{e}^{-16}$   & $\textrm{e}^{-12}$ & (a)&  6& 2  & $\textrm{e}^{-12}$& $\textrm{e}^{-4}$& (e)& 6& 4&  $\textrm{e}^{-12}$   & $\textrm{e}^{-8}$& (i)  \\
2& -2& 2& 2&               6& 6&   $\textrm{e}^{-12}$   & $\textrm{e}^{-12}$& (a) & 2& 2&   $\textrm{e}^{-4}$& $\textrm{e}^{-4}$& (e)& 4& 4&  $\textrm{e}^{-8}$& $\textrm{e}^{-8}$ & (i)\\
2& 2& -2& 2&            6& 6&      $\textrm{e}^{-12}$   & $\textrm{e}^{-12}$ & (a)&  2& 2&   $\textrm{e}^{-4}$& $\textrm{e}^{-4}$& (e)& 4& 4 & $\textrm{e}^{-8}$& $\textrm{e}^{-8}$& (i) \\
2& 2& 2& -2&         6& 6&    $\textrm{e}^{-12}$   & $\textrm{e}^{-12}$& (b) &  2& 2&    $\textrm{e}^{-4}$& $\textrm{e}^{-4}$& (f)& 4& 4 & $\textrm{e}^{-8}$& $\textrm{e}^{-8}$& (j)\\
-2& -2& 2& 2&           6& 2&   $\textrm{e}^{-12}$   & $\textrm{e}^{-4}$ & (b)&  2& -2&    $\textrm{e}^{-4}$& $C_1$& (f)&  4& 0 & $\textrm{e}^{-8}$& $h^{-\gamma}$& (j)\\
-2& 2& -2& 2&           4& 2&  $\textrm{e}^{-8}$   & $\textrm{e}^{-4}$& (b) & 0& -2&    $ C_2$& $C_1$& (f)&  2& 0& $\textrm{e}^{-4}$& $h^{-\gamma}$& (j)\\
-2& 2& 2& -2&           4& 2&   $\textrm{e}^{-8}$   & $\textrm{e}^{-4}$& (b) & 0& -2&   $ C_2$& $C_1$& (f)& 2& 0& $\textrm{e}^{-4}$& $h^{-\gamma}$& (j) \\
2& -2& -2& 2&         4& 2& $\textrm{e}^{-8}$   & $\textrm{e}^{-4}$ & (c) &  0 & -2&   $ C_2 $& $C_1$& (g)&   2& 0& $\textrm{e}^{-4}$& $h^{-\gamma}$& (k) \\
2& -2& 2& -2&      2& 2&   $\textrm{e}^{-4}$   & $\textrm{e}^{-4}$& (c)& -2& -2&   $C_3$& $C_1$& (g)& 0& 0& $h^{-\xi}$& $h^{-\gamma}$& (k) \\
2& 2& -2& -2&           2& 2&    $\textrm{e}^{-4}$   & $\textrm{e}^{-4}$& (c) &  -2& -2&     $C_3$& $C_1$& (g)& 0& 0 & $h^{-\xi}$& $h^{-\gamma}$& (k) \\
-2& -2& -2& 2&           4& -2&     $\textrm{e}^{-8}$   & $C_1$ & (c) & 0& -6&    $C_2$& $C_1$& (g)&  2& -4& $\textrm{e}^{-4}$& $C_1$& (k)  \\
-2& -2& 2& -2&               2& -2&        $\textrm{e}^{-4}$   & $C_1$& (d)& -2& -6&  $C_3$& $C_1$& (h)&  0& -4 & $h^{-\xi}$& $C_1$ & (l)\\
-2& 2& -2& -2&           2& -2&        $\textrm{e}^{-4}$   & $C_1$& (d)& -4& -6&     $C_3$& $C_1$& (h)& -2& -4& $h^{-\xi}$& $C_1$& (l) \\
2& -2& -2& -2&           2& -2&      $\textrm{e}^{-4}$   & $C_1$& (d)& -6& -6&    $C_3$& $C_1$& (h)&  -4& -4 & $h^{-\xi}$& $C_1$& (l) \\
-2& -2& -2& -2&          2 & -6 &      $\textrm{e}^{-4}$   & $C_1$& (d) &  -4 & -10 &   $C_3$ & $C_1$& (h) & -2 & -8 & $h^{-\xi}$ & $C_1$ & (l) \\
\hline \hline
\end{tabular}
\label{maximumenergyforM}
\end{table}
%\end{landscape}

%\section{Figures}

\begin{figure}[h]
\centering
\includegraphics[width=\textwidth]{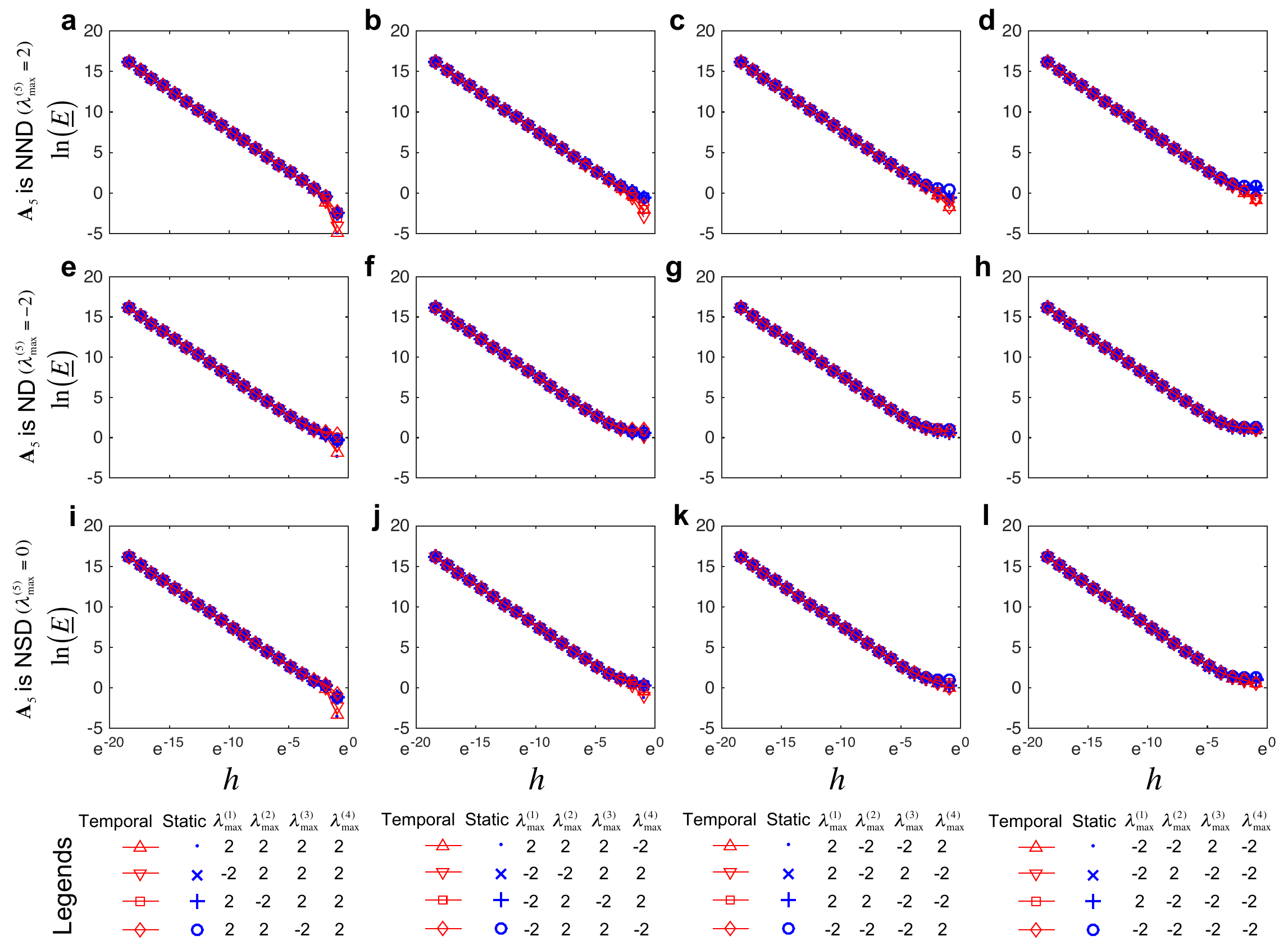}
\caption{
Numerical results for the scaling behavior of $\underline{E}$ for $5$ snapshots when $h$ is small.
For each panel, all related parameters are shown in Table~\ref{minimumenergyforM}, and the numerical calculations are in excellent agreement with the theoretical results shown in the same table.
}
\label{figsi_min1}
\end{figure}

\begin{figure}
\centering
\includegraphics[width=\textwidth]{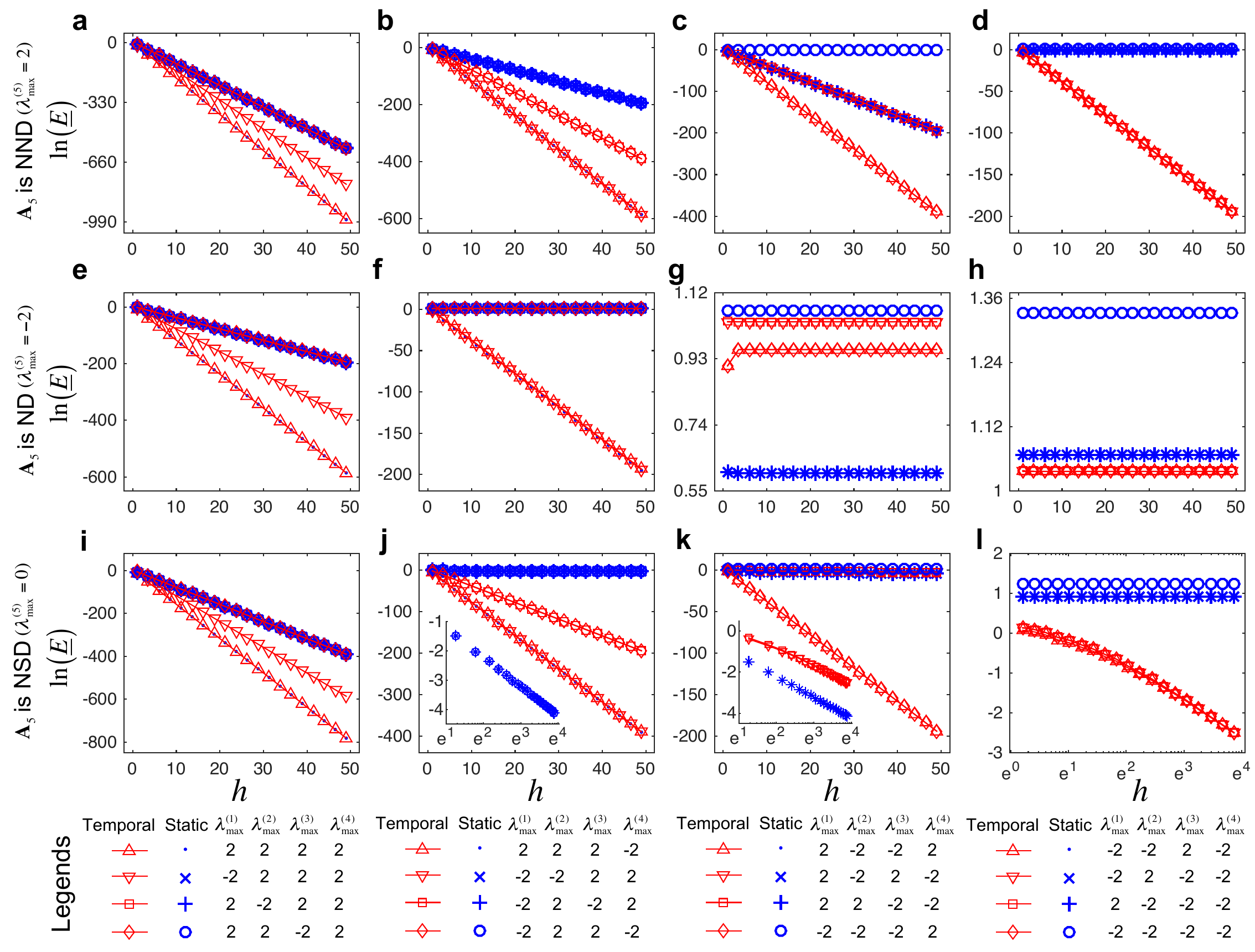}
\caption{
Numerical results for the scaling behavior of $\underline{E}$ for $5$ snapshots when $h$ is large.
In panels (j) and (k), insets give a clearer representation of $\underline{E}$ for a short interval.
For each panel, all related parameters are shown in Table~\ref{minimumenergyforM}, and the numerical calculations are in excellent agreement with the theoretical results shown in the same table.
}
\label{figsi_min2}
\end{figure}

\begin{figure}
\centering
\includegraphics[width=\textwidth]{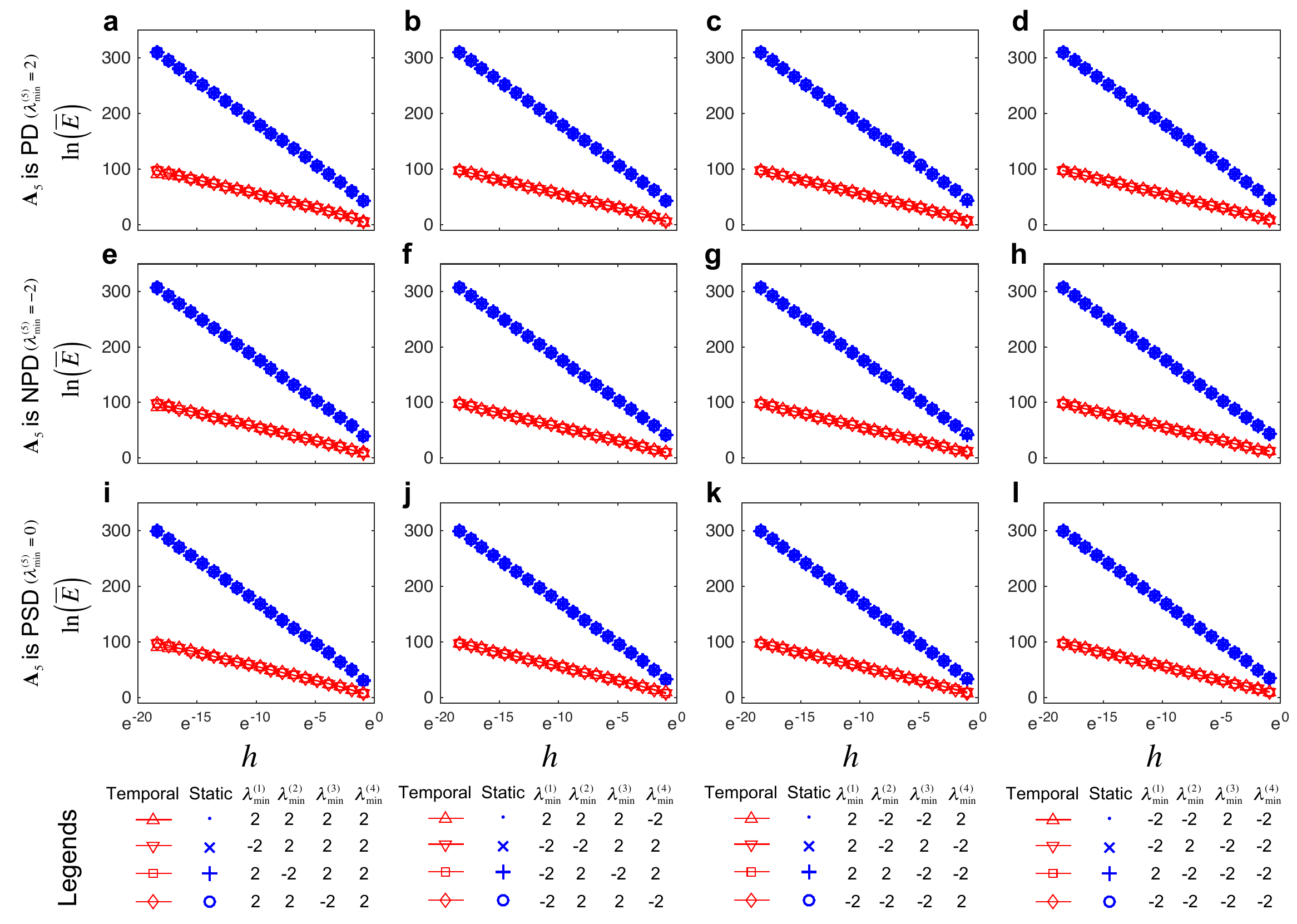}
\caption{Numerical results for the scaling behavior of $\overline{E}$ for $5$ snapshots when $h$ is small.
For each panel, all related parameters are shown in Table~\ref{maximumenergyforM}, and the numerical calculations are in excellent agreement with the theoretical results shown in the same table.
}
\label{figsi_max1}
\end{figure}

\begin{figure}
\centering
\includegraphics[width=\textwidth]{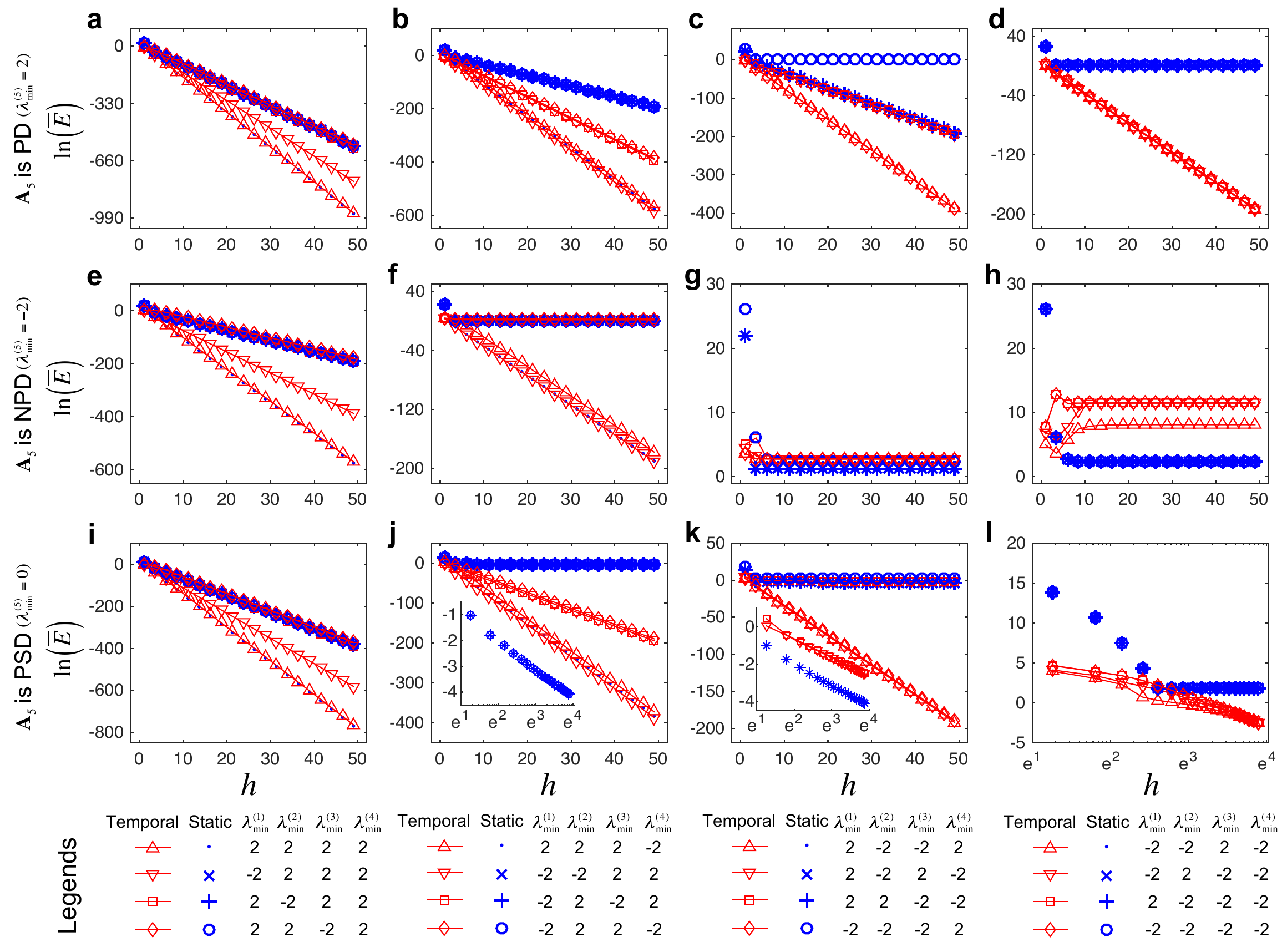}
\caption{
Numerical results for the scaling behavior of $\overline{E}$ for $5$ snapshots when $h$ is large.
In panels (j) and (k), insets give a clearer representation of $\overline{E}$ over a short interval.
For each panel, all the related parameters are shown in Table~\ref{maximumenergyforM}, and numerical calculations are in excellent agreement with the theoretical results shown in the same table.
}
\label{figsi_max2}
\end{figure}

\end{document}